\documentclass[10pt, conference, a4paper]{IEEEtran}
\IEEEoverridecommandlockouts	
\IEEEpubid{\makebox[\columnwidth]{979-8-3503-1258-4/23/\$31.00~\copyright~2023 IEEE \hfill} \hspace{\columnsep}\makebox[\columnwidth]{ }}
\usepackage{cite}
\usepackage{amsmath,amssymb,amsfonts}
\usepackage{algorithmic}
\usepackage{graphicx}
\usepackage{textcomp}
\usepackage{xcolor}
\usepackage[acronym]{glossaries}
\usepackage{eurosym}
\usepackage{tikz}
\usetikzlibrary{shapes}
\usetikzlibrary {positioning}
\usepackage{url}
\usepackage{placeins}
\usepackage{hyperref}
\usepackage{mdframed}
\newmdenv[
  topline=false,
  bottomline=false,
  rightline=false,
  skipabove=\topsep,
  skipbelow=\topsep,
  leftmargin=0pt,
  rightmargin=-10pt,
  innertopmargin=0pt,
  innerbottommargin=0pt
]{siderules}

\def\BibTeX{{\rm B\kern-.05em{\sc i\kern-.025em b}\kern-.08em
    T\kern-.1667em\lower.7ex\hbox{E}\kern-.125emX}}

\begin{document}
\newacronym{EC}{EC}{European Commission}
\newacronym[plural=RFNBOs,firstplural=renewable fuels of non-biological origin (RFNBOs)]{RFNBO}{RFNBO}{renewable fuel of non-biological origin}
\newacronym{PPA}{PPA}{Power Purchase Agreement}
\newacronym{IRA}{IRA}{Inflation Reduction Act}
\newacronym[plural=GOs,firstplural=guarantees of origin (GOs)]{GO}{GO}{guarantee of origin}
\newacronym{FLH}{FLH}{full-load hours}
\newacronym{P2X}{P2X}{Power-to-X}
\newacronym{RES}{RES}{renewable energy sources}
\newacronym{MILP}{MILP}{mixed-integer linear problem}
\newacronym{ASU}{ASU}{air-separation unit}
\title{Renewable fuel regulation: Implications for e-fuel production infrastructure in energy hubs
\thanks{The authors would like to acknowledge financial support from the SuperP2G project in the European Union’s Horizon 2020 research and innovation programme under grant agreement No 775970, the H2Elements project under the Marie Skłodowska–Curie grant agreement No 899987, and under the Nordic Energy Research's mobility and network program NordNET under grant agreement No. 119646.}
}

\author{\IEEEauthorblockN{Lissy Langer\IEEEauthorrefmark{1}, Ioannis Kountouris, Rasmus Bramstoft, Marie M\"unster and Dogan Keles}
\IEEEauthorblockA{Technical University of Denmark (DTU), Section for Energy Economics and Modeling, \\Produktionstorvet, Bld. 424, 2800 Kgs. Lyngby, Denmark} \IEEEauthorblockA{\IEEEauthorrefmark{1}Corresponding author: lilang@dtu.dk}
}

\maketitle
\pagestyle{plain}

\begin{abstract}
\Glspl{RFNBO} are needed to decarbonize hard-to-electrify sectors which rely on liquid or gaseous fuels, such as long-haul shipping. The EU's Delegated Act on \Glspl{RFNBO} defines \textit{renewable} hydrogen by considering rules on additionality as well as temporal and geographical correlation of the electricity used. For a Danish case study, we examine the impact on the capacity expansion problem of an energy hub producing renewable hydrogen, e-methanol, and e-ammonia using a mixed-integer linear problem formulation. We analyze the investments in production capacity, storage assets, and \gls{PPA} volume under different fuel price assumptions for 2030. We find that e-methanol (combined with limited storage to secure hydrogen supply to the synthesizer) provides the best business case with a PPA volume based on the maximum allowed electrolyzer size. 
\end{abstract}


\section{Introduction}
In its REPowerEU plan published in May 2022, the \gls{EC} sets out to accelerate the clean energy transition and reduce the EU's dependency on fossil fuel imports \cite{european_commission_commission_2023-1}. The aim is to reach 10~million tonnes of domestic renewable hydrogen production and 10~million tonnes of imports by 2030 to replace natural gas, coal, and oil in hard-to-abate industry and transport sectors \cite{european_commission_repowereu_2022}. Nevertheless, the use of \textit{renewable} electricity is essential for hydrogen production, as hydrogen from fossil electricity is more emission-intensive than hydrogen from natural gas through steam methane reforming \cite{european_commission_commission_2023-1}. The \gls{EC} expects that around 500--550\,TWh or \euro 200--300 billion of additional renewable energy production will be needed to fulfil the EU targets \cite{european_commission_commission_2023}. 

In Feb 2023, two Delegated Acts were published to provide regulatory certainty on what constitutes \textit{renewable} hydrogen and fuels in the EU. These clarifications aim to ensure that net decarbonisation is achieved by using only \textit{additional} renewable electricity produced at the same time and in the same region (see Section~\ref{sec:eu-reg}). The Delegated Acts set guidelines for infrastructure investments, state aid, hydrogen imports, and for the proposed renewable consumption targets for industry and the transport sector of the Fit for 55 package \cite{european_commission_european_2021}.

\subsection{EU Delegated Act on Renewable Hydrogen and Fuels} \label{sec:eu-reg}
In Dec 2018, the Renewable Energy Directive (RED\,II) introduced a renewable fuel quota of 14\% for fuel suppliers in road and rail transport by 2030. In addition, it requested the \gls{EC} to develop a Union methodology for the electricity used to produce \glspl{RFNBO} to be considered as \textit{fully renewable} \cite{european_commission_directive_2018}. In Feb 2023, the Delegated Act on Renewable Hydrogen \cite{european_commission_commission_2023-1} defined three ways to source fully renewable electricity: 
\begin{enumerate}
    \item \textbf{using on-site RES} no older than 36 months at the start of the \gls{RFNBO} production (§\,3)
    \item\textbf{using a \gls{PPA}} following rules on additionality as well as temporal and geographical correlation (§\,5)
    \item \textbf{using the grid} given a high share of RES in the past calendar year (§\,4.1), or at the time of consumption leading to downward dispatching (§\,4.3), or a low day-ahead or CO$_{2_{eq}}$ allowance price (§\,6).
\end{enumerate}

In addition, the \gls{EC} added an exception for low-carbon countries (mainly with a high share of nuclear power)\footnote{Separate regulation for low-carbon hydrogen using electricity produced from nuclear power might be introduced earliest by the end of 2024. Blue hydrogen (from natural gas with CCS) will most likely not be subsidized.}, waiving the additionality rule on renewable \gls{PPA}s (§\,4.2).

\textbf{The Additionality Rule} aims to introduce additional renewable energy capacity so as not to divert from regular electricity demand. A \gls{PPA} can include renewable power plants not built more than 36 months before the start of \gls{RFNBO} production (§\,5.a). The rule disallows stacking financial support and \glspl{GO} for the fuel produced and electricity used (§\,5.b). Until 2028, electrolyzers can still sign \gls{PPA}s with existing renewable power plants not to delay investments.

\textbf{The Temporal Correlation Rule} aims to increase the share of \gls{RES} and storage assets with a higher system value by matching supply and demand in a given hour. Until 2030, only monthly matching is enforced to allow for constant delivery without hydrogen infrastructure and storage yet in place (member states can introduce hourly matching as of July 2027 upon notification).

\textbf{The Geographical Correlation Rule} aims to achieve \textit{deliverability} of the introduced \gls{RES} generation to limit the burden on the transmission infrastructure. A region also includes adjacent offshore wind zones as well as interconnected zones with a higher day-ahead price to reduce network congestion.

Simultaneously, the US Treasury Department is in the process of defining an equivalent US methodology, most likely also combining rules on additionality as well as temporal and geographical correlation. The financial gains are substantial, as the US \gls{IRA} subsidizes hydrogen production with up to \$3 per kg\textsubscript{H\textsubscript{2}} based on emission intensity. In addition, during the revision of the Greenhouse Gas Protocol's scope 2 emission standards, similar measures are being discussed \cite{world_resources_institute_wri_ghg_2023}.

\subsection{EU Renewable Hydrogen Support}
In Feb 2023, the \gls{EC} published its Green Deal Industrial Plan, claiming that it will be more cost-effective, fast, and administratively light than the US \gls{IRA} \cite{european_commission_green_2023}. It supports the scale-up of EU manufacturing capacity for net-zero technologies and products.
The EU Innovation Fund will---through competitive bids---cover the funding gap between renewable hydrogen and fossil fuels by introducing a fixed premium for each kg of renewable hydrogen produced over a period of 10 years. The first auction of the European Hydrogen Bank is to take place in the autumn/end of 2023 with an indicative budget of \euro 800 million and can, using an auction-as-a-service scheme, be extended with additional national funding \cite{european_commission_commission_2023}. Terms and conditions of the initial carbon contracts for difference (CCFD) format are to be published in May and finalized in June 2023.

\subsection{EU Renewable Hydrogen Production}
The \gls{EC} aims at 6000\,MW\textsubscript{e} of electrolyzers by the end of 2025. Yet, the market is still emerging since most of the current 160\,MW\textsubscript{e} of electrolyzer installations are demonstration plants with the largest plant currently under construction being 20\,MW\textsubscript{e} \cite{european_commission_questions_2023}. The certainty on regulation and funding might lead to additional investments on the supply side. On the demand side, however, e.g.,  in the shipping sector, technology choices, future demand, and prices are still undetermined. The question remains how the gap between willingness-to-pay and production costs can be bridged in highly competitive yet hard-to-abate sectors. In this regard, we investigate the production of e-methanol and e-ammonia, both considered relevant for decarbonizing long-haul shipping.

 We investigate the impact of the proposed EU regulation on the investment decisions of an energy hub producing renewable hydrogen (H\textsubscript{2}), e-methanol (MeOH), and e-ammonia (NH\textsubscript{3}). We combine the regulatory with the operational constraints and analyze varying flexibility requirements, such as additional investments in local battery and/or hydrogen storage assets. In particular, we evaluate the impact of:
 \begin{itemize}
     \item renewable fuel prices,
     \item level of capital expenditures, and
     \item level of operational details modelled
 \end{itemize}
 on the need for flexibility (mainly represented by storage assets), the \gls{FLH}, as well as the production capacities, fuels produced, and overall cost of the system.

\section{Related Work}
For the Western US,  Ricks et al. (2023) investigate the impact of the \gls{IRA} hydrogen subsidies on the system, analyzing generation and storage capacities, cost, and emissions evaluating varying temporal correlation rules \cite{ricks_minimizing_2023}. In a European context, two working papers investigate regulatory aspects in a German case study \cite{brauer_green_2022} and a number of EU countries considering the flexibility of hydrogen demand and different types of hydrogen storage assets \cite{zeyen_hourly_2022}. However, the detailed ruling of the Delegated Act was not considered yet. In addition, the studies only consider the production and demand of hydrogen and not its derivatives. On the other hand, capacity expansion models of energy hubs predominantly focus on technical operation on a shorter time scale or policy evaluation on a macroeconomic scale with less technical detail.

\section{Methodology}
The energy hub modelled is inspired by the industrial park \href{www.greenlab.dk/}{GreenLab Skive (GLS)} in North-West Denmark. The park aims to create a synergetic network for its partners on site. It plans local \gls{RES} as well as producing hydrogen and potential derivatives. Therefore, we use this representative case to illustrate the respective investment decisions considering the renewable hydrogen regulation.

\begin{figure}[ht]
\vspace{-10pt}
\centering
\includegraphics[scale=0.40, trim = {120, 150, 100, 70}, clip]{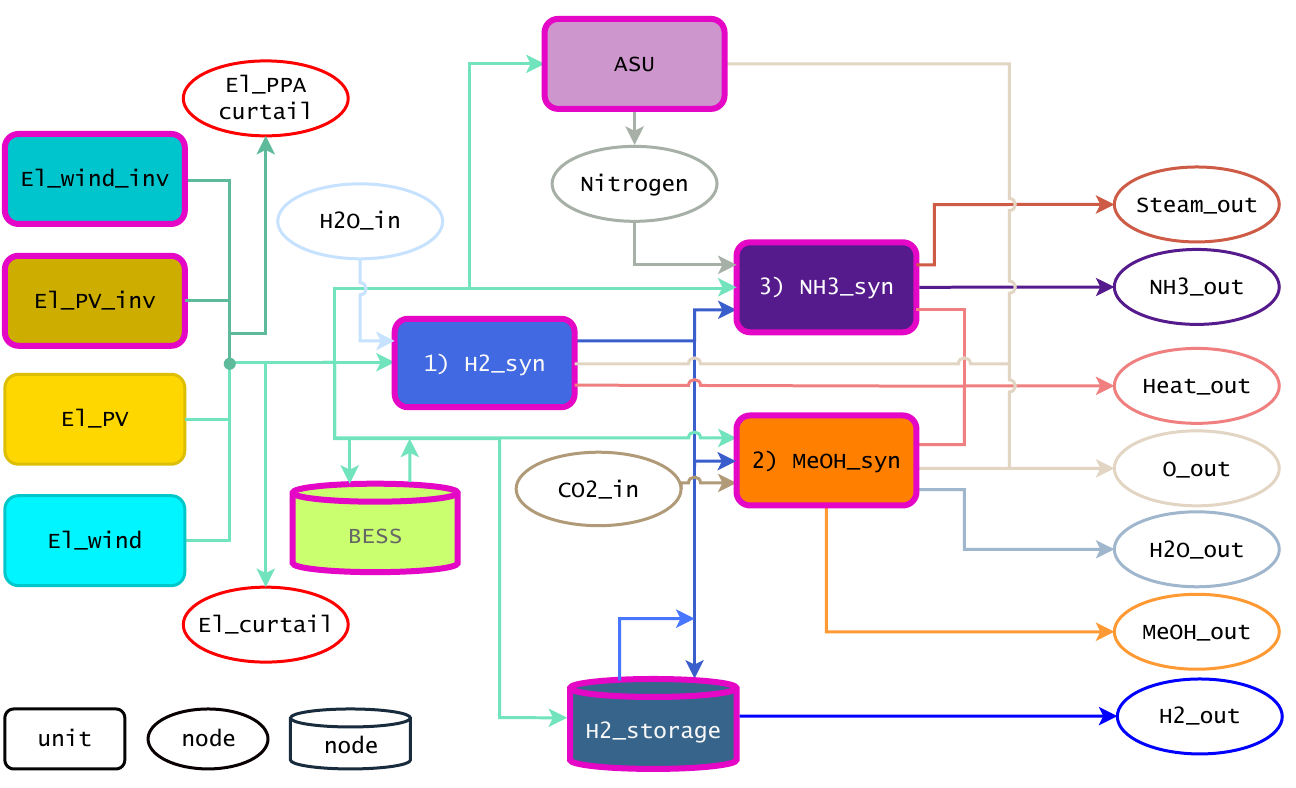} 
\caption{GreenLab Skive energy park technology flow chart (magenta frames indicate investment decisions) adapted based on \cite{kountouris_power--x_2023}.}
\label{fig:skive-flow-chart}
\end{figure}

Fig.~\ref{fig:skive-flow-chart} illustrates the technical assets modelled for GLS. Investment assets are indicated with a magenta frame. We analyze three distinct setups of the energy hub: a) only H\textsubscript{2}, b) H\textsubscript{2} and MeOH, or c) H\textsubscript{2} and NH\textsubscript{3} production. In addition, we investigate the sensitivity of the results for different levels of fuel prices and capital expenditures. The model developed is based on the EnerHub2X model that was used to investigate the operational patterns of the GLS energy hub \cite{kountouris_power--x_2023}. The model was expanded to consider investments.

\subsection{Modeling Framework}
While the initial model used GAMS, we use the \textit{SpineOpt} modelling framework, which allows for linear and \gls{MILP} formulation within the same framework and flexible scenario generation and temporal resolution \cite{ihlemann_spineopt_2022}. The model is constructed by defining unit and node objects, as well as their relationships and respective parameter values, using the underlying constraints in a flexible generalized manner. The data flow is managed using \textit{Spine Toolbox}, which can also be integrated with other modelling frameworks \cite{kiviluoma_spine_2022}. In the following, we briefly explain how the entities are modelled. The detailed data can be found in Appendix \ref{sec:data-input} and on \href{https://github.com/lilanger/Enerhub2X.spineopt}{GitHub}.
 
\subsection{Units}
As indicated in Fig.~\ref{fig:skive-flow-chart}, units are used to represent converters such as the synthesizers and RES assets. Units can be invested in and can have multiple in- and outputs (for data see Tab.\ref{tab:unit_param}).

\paragraph{Investment Decisions and Unit Commitment Types}Capacity investment decisions can be continuous or integer units of a specific size---introducing a \gls{MILP}. Using a \gls{MILP} formulation, one can also introduce unit commitment constraints such as minimum loads or ramping constraints.

The \gls{MILP} model includes a minimum operating level of 20\% and an (optimistic) cold-start ramp-up time of two hours for all \gls{P2X} synthesizers to make operations more realistic. The impact will be discussed briefly in Sec.\ref{sec:sensitivity}. The LP version of the model does not include unit commitment, and all units can be invested in continuously.

\begin{siderules}
\textbf{For example}, the investment in  the \gls{ASU} is continuous and does not include unit commitment. The electrolyzer, on the other hand, is represented using distinct stacks of 450\,kW\textsubscript{e} each. Then, unit commitment applies to each separate stack unit. Finally, the \gls{P2X} synthesizers are represented by units with distinct capacities ranging from 1 to 10 tons of fuel output per hour. Then, unit commitment applies to the whole unit. 

\begin{center}
    \vspace{10pt}
    \scalebox{0.9}{
    \begin{tabular}{ | l | c | c | c |}
       \hline
       \textbf{Unit} & \textbf{Candidate} & \textbf{Unit} & \textbf{Investment}\\
       \textbf{} & \textbf{Units} & \textbf{Capacity} & \textbf{Type}\\ \hline
       ASU & 10  & 1 tons\textsubscript{N}/h & continuous\\ 
       H\textsubscript{2}\_syn & 222 & 8.1 kg\textsubscript{H\textsubscript{2}}/h & integer\\ 
       NH\textsubscript{3}\_syn & 1 & 1--10 tons\textsubscript{NH\textsubscript{3}}/h & binary\\ \hline
    \end{tabular}
    }
\end{center}
\end{siderules}

\paragraph{Input-Output Ratios} are used to fix the relationship between in- and outputs. They can be defined between two inputs, two outputs, or an in- and an output.

\begin{siderules}
\textbf{For example}, the electrolyzer takes water (\textit{H\textsubscript{2}O\textsubscript{in}}) and electricity as input and produces hydrogen (\textit{H\textsubscript{2}}), oxygen (\textit{O\_out}), and heat (\textit{Heat\_out}). The fixed input-output ratio for electricity and water are 53.6 and 9.99, respectively, so for one ton of hydrogen 53.6\,MWh of power and 9.99 tons of water are needed. At the same time, the output-output ratios define that per ton of hydrogen 9.76\,MWh heat and 4.96 tons of oxygen are produced.

\begin{center}
    \begin{tikzpicture}[
      unit/.style={rectangle, rounded corners, minimum width=1.5cm, minimum height=2cm,text centered, draw=black, fill=white, ultra thick},
      node/.style={ellipse, minimum width=1.8cm, minimum height=.5cm, text centered, draw=black, fill=white, ultra thick},
     arrow/.style={thick,->,>=stealth}]
    
        \node (a1) [node, yshift=.5cm, draw={rgb,255:red,198; green,226; blue,255}] {H\textsubscript{2}O\textsubscript{in}};
        \node (a2) [node, below=.1cm of a1, draw={rgb,255:red,120; green,205; blue,170}] {Group\textsubscript{El}};
        \node (b)  [unit, xshift=3cm, draw={rgb,255:red,65; green,105; blue,225}] {H2\textsubscript{syn}};
        \node (c1) [node, xshift=6cm, yshift=.9cm, draw={rgb,255:red,58; green,95; blue,205}] {H\textsubscript{2}};
        \node (c2) [node, below=.1cm of c1, draw={rgb,255:red,240; green,128; blue,128}] {Heat\textsubscript{out}};
        \node (c3) [node, below=.1cm of c2, draw={rgb,255:red,250; green,235; blue,215}] {O\textsubscript{out}};
    
        \foreach \x [count = \xi] in {9.99, 53.6}
        \draw [arrow,->] (a\xi.east) -- (b.west |- a\xi.east) node[midway,above] {\x};
        \foreach \y [count = \yi] in {1, 9.76, 4.96}
        \draw [arrow,->] (b.east |- c\yi.west) -- (c\yi.west) node[midway,above] {\y};
    \end{tikzpicture}
\end{center}
\end{siderules}

\paragraph{Unit Availability and Flow Variables}
The variability of the \gls{RES} units is modelled using a \texttt{unit\_availabilty\_factor}. For an existing unit, the factor is multiplied with the \texttt{unit\_capacity}, the following (unit commitment) variables are bounded accordingly:\\ $\geq$ \texttt{units\_available} $\geq$  \texttt{units\_on} $\geq$ \texttt{unit\_flow}.  For investments, the availability is determined by the \texttt{units\_invested} in. 

\begin{siderules}
\textbf{For example}, PV can be invested in to approximate a \gls{PPA} connection \cite{zeyen_hourly_2022}. The following parameters are defined:\\

\begin{center}
    \scalebox{0.65}{
        \begin{tabular}{ | l | c | c | c |}
           \hline
           \textbf{Class} & \textbf{Object} & \textbf{Parameter} & \textbf{Value} \\ \hline
            unit & PV & \texttt{candidate\_units} & 1000   \\
            unit & PV & \texttt{number\_of\_units}          & 0    \\
            unit & PV & \texttt{unit\_availability\_factor}      & time series \\
            unit & PV & \texttt{unit\_investment\_cost}          & 41,259    \\
            unit & PV & \texttt{unit\_investment\_variable\_type} & continuous \\
            unit\_to\_node & PV,El\_PPA & \texttt{unit\_capacity} & 1 \\ \hline
        \end{tabular}
    }
\end{center}

Accordingly, the system could invest continuously in PV assets up to 1000\,MW with an availability defined by the hourly \texttt{unit\_availability\_factor} and with annualized investments of \euro 41,259 per MW. The unit flow from the PV asset to the PPA electricity node is then restricted by the following equations: 
\end{siderules}
    
The investment and retirement per investment period (e.g., years) define the \textit{available} investments restricted by the \texttt{candidate\_units}. This is the case for all investments in units and nodes (where variables are called \texttt{storages\_invested}... instead of \texttt{units\_invested}...).

    \vspace{-12pt}
    \begin{equation}\label{eq:invest_cand}
        \begin{split}
            \mathtt{units\_invested\_available_t} \leq  \mathtt{candidate\_units} \\
            \hspace{2em} \forall t \in \text{investment periods}
        \end{split}
    \end{equation}

    \vspace{-12pt}
    \begin{equation} 
        \begin{split} \label{eq:invest_avail}
            \mathtt{units\_invested\_available_t} = \\
            \mathtt{units\_invested\_available_{t-1}}  + \mathtt{units\_invested_t} \\
            - \mathtt{units\_mothballed_t} \hspace{3em} \forall t \in \text{investment periods} 
        \end{split}
    \end{equation}
       
    The \texttt{units\_invested\_available} are the upper bound for the \texttt{units\_available} per period (e.g., hours), taking into account initial units and the availability factor.

    \vspace{-12pt}
    \begin{equation}
        \begin{split}
            \mathtt{units\_available_h} \leq \\
            \mathtt{units\_availability\_factor_h}  \\
            \times (\mathtt{number\_of\_units} + \mathtt{units\_invested\_available_t} ) \\
            \forall h \subset t \in \text{investment periods} 
        \end{split}
    \end{equation}

    The \texttt{units\_available} are the upper bound for the \texttt{units\_on} per period. 

    \vspace{-10pt}
    \begin{equation} \label{eq:invest_on}
        \mathtt{units\_on_h} \leq \mathtt{unit\_available_h} \hspace{2em} \forall h 
    \end{equation}

    The \texttt{units\_on} times the \texttt{unit\_capacity} are the upper bound for the \texttt{unit\_flow} per period. 
    
    \vspace{-18pt}
    \begin{equation} \label{eq:invest_flow}
        \mathtt{unit\_flow_h} \leq \mathtt{units\_on_h} \times \mathtt{unit\_capacity} \hspace{1em} \forall h
    \end{equation}
    \vspace{-18pt}

Since the availability factors do only set an upper bound and do not fix the \texttt{units\_available} as the \texttt{units\_on}, one has to carefully model \gls{RES} curtailment.

\subsection{Nodes}
There are four different kinds of nodes illustrated in Fig.~\ref{fig:skive-flow-chart}: sources, sinks, intermediate products, and storage assets (data see Tab.\ref{tab:node_param}).

\paragraph{Nodal Balance Sense}
The first three types have distinct nodal balance senses.
\begin{center}
    \scalebox{0.9}{
    \begin{tabular}{ | l | c | c | }
       \hline
       \textbf{Type} & \textbf{Example Node} & \textbf{Nodal Balance Sense}\\ \hline
       Sources & \texttt{CO2\_in} & \texttt{<=}  \\ 
       Sinks & \texttt{H2\_out} & \texttt{>=}  \\ 
       Products & \texttt{H2} & \texttt{==}  \\ \hline
    \end{tabular}
    }
\end{center}

 The sources create a product (e.g., CO\textsubscript{2}) that becomes the input to units, inducing a \texttt{fuel\_cost}. Both CO\textsubscript{2} and H\textsubscript{2}O are assumed to be infinitely available.  The sinks consume a product (e.g., H\textsubscript{2}), inducing a negative \texttt{fuel\_cost} as the fuels are the main products sold. Intermediate products need to be balanced, e.g., the hydrogen from the electrolyser must either be stored, compressed, or used in a \gls{P2X} unit.

\paragraph{Storage Assets}
Nodes can work as storage assets by setting \texttt{has\_state}. Subsequently, storage equality constraints apply, e.g. for BESS, these parameters are set:

\begin{center}
    \scalebox{0.9}{
        \begin{tabular}{ | l | c | c | c |}
           \hline
           \textbf{Class} & \textbf{Object} & \textbf{Parameter} & \textbf{Value} \\ \hline
            node & BESS & \texttt{has\_state}        & True        \\
            node & BESS & \texttt{fix\_node\_state}  & time series \\
            node,t & BESS,t  & \texttt{cyclic\_condition} & True\\
            node & BESS & \texttt{frac\_state\_loss} & 4e-05       \\
            node & BESS & \texttt{node\_state\_cap}  & 1  \\ \hline
        \end{tabular}
    }
\end{center}

The initial state of charge can be defined using \texttt{fix\_node\_state}, with the \texttt{cyclic\_condition} as True this state also has to be reached again at the end of the planning horizon. The \texttt{frac\_stat\_loss} defines a percentage loss per time period and \texttt{node\_state\_cap} the upper limit of the state of charge. Considering storage investments, this value is multiplied by \texttt{storages\_invested\_available} similar to Eq.~\ref{eq:invest_on} to restrict the \texttt{node\_state} at a given time.

\paragraph{Groups} These sum commodities of the same kind, e.g., to define \texttt{unit\_capacities} and \texttt{fixed\_ratios} over the sum of a group. Groups can, however, not be used to define common parameters such as \texttt{fixed\_ratios} or \texttt{fuel\_cost} for assets of the same kind.

\begin{siderules}
\textbf{For example}, the different sources of renewable electricity are combined in the \texttt{Group\_El} that can be used as input, e.g., to the electrolyzer with a \texttt{fix\_ratio\_in\_out\_unit\_flow} of 53.6\,MWh per ton of hydrogen.

\begin{center}
    \begin{tikzpicture}[
      unit/.style={rectangle, rounded corners, minimum width=1.5cm, minimum height=3cm,text centered, draw=black, fill=white, ultra thick},
      node/.style={ellipse, minimum width=1.8cm, minimum height=.5cm, text centered, draw=black, fill=white, ultra thick},
     arrow/.style={thick,->,>=stealth}]
    
        \node (a1) [node, yshift=.9cm, draw={rgb,255:red,127; green,255; blue,212}] {El\textsubscript{RES}};
        \node (a2) [node, below=.1cm of a1, draw={rgb,255:red,102; green,205; blue,170}] {El\textsubscript{PPA}};
        \node (a3) [node, below=.1cm of a2, draw={rgb,255:red,128; green,128; blue,0}] {El\textsubscript{BESS}};
        \node (b)  [node, minimum height=3cm, xshift=3cm, draw={rgb,255:red,120; green,205; blue,170}] {Group\textsubscript{El}};
        \node (c)  [unit, xshift=6cm, draw={rgb,255:red,65; green,105; blue,225}] {H2\textsubscript{syn}};

        \foreach \x [count = \xi] in {1, 1, 1}
        \draw [arrow,->] (a\xi.east) -- (b.west |- a\xi.east) node[midway,above] {\x};
        \draw [arrow,->] (b.east |- c.west) -- (c.west) node[midway,above] {53.6};
    \end{tikzpicture}
\end{center}
\end{siderules}

\section{Results}
\begin{figure*}[ht]
\centering
\includegraphics[width=\textwidth, trim = {12, 12, 11, 15}, clip]{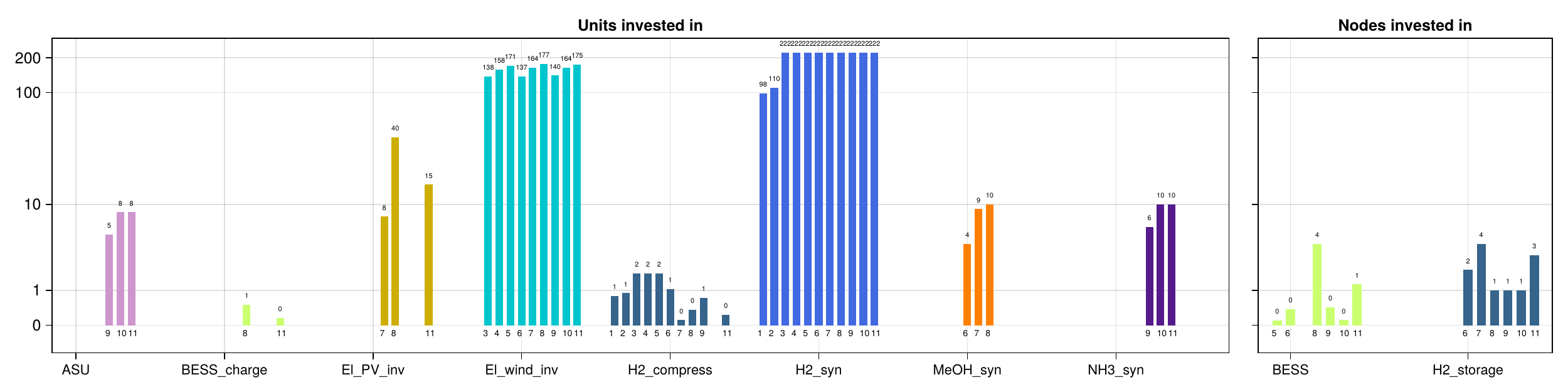} 
\caption{GreenLab Skive energy park investment decisions based on scenarios in Table~\ref{tab:scenario_description} (non-indicated values are zero).}
\label{fig:investresults}
\end{figure*}

\begin{figure*}[ht]
\centering
\includegraphics[width=\textwidth, trim = {15, 15, 15, 15}, clip]{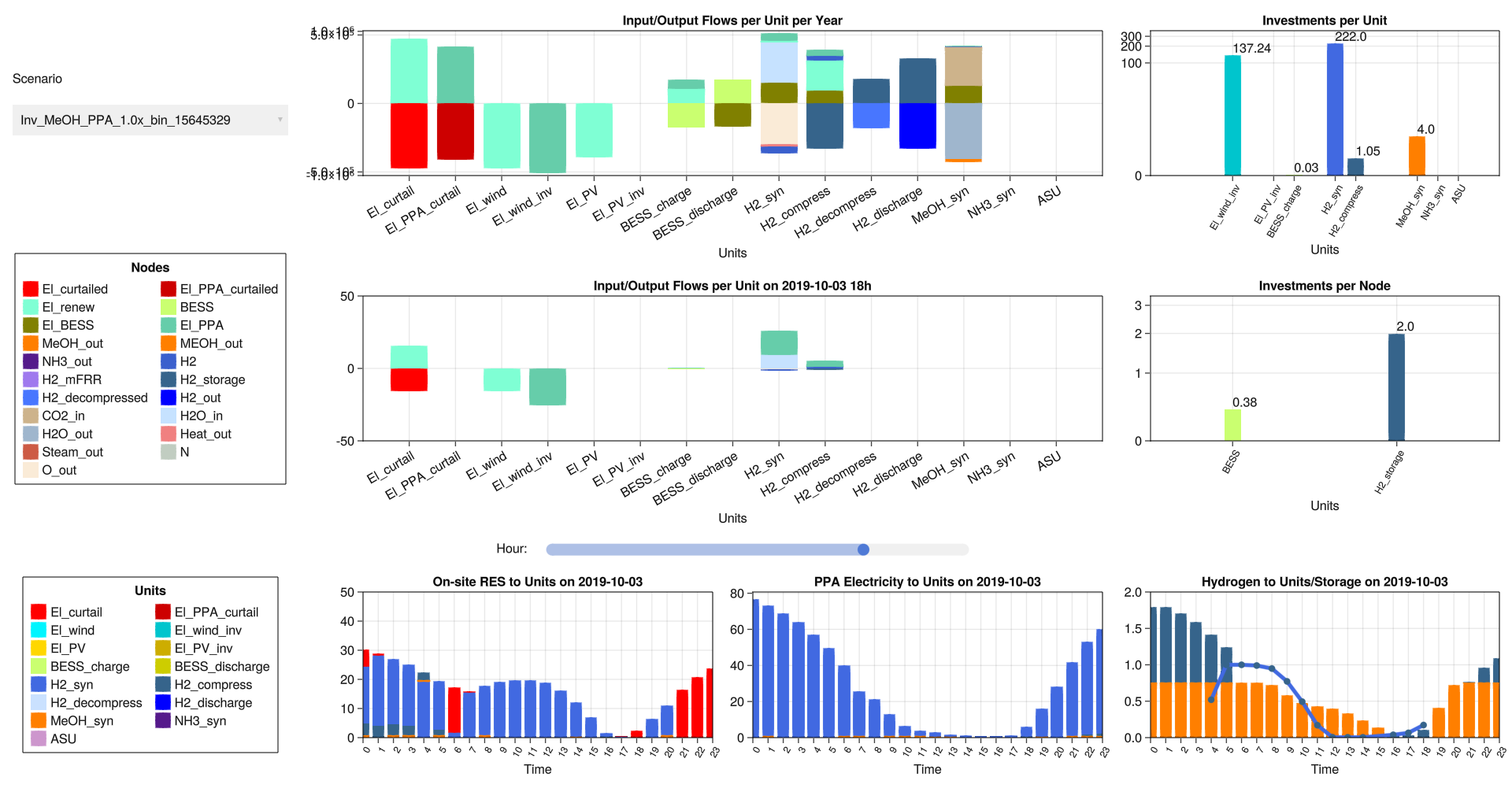} 
\caption{GreenLab Skive energy park investment and illustrative operational decisions of scenario S6 in Table~\ref{tab:scenario_description} (non-indicated values are zero).}
\label{fig:db_s6}
\end{figure*}

\begin{figure*}[ht]
\centering
\includegraphics[width=\textwidth, trim = {15, 15, 15, 15}, clip]{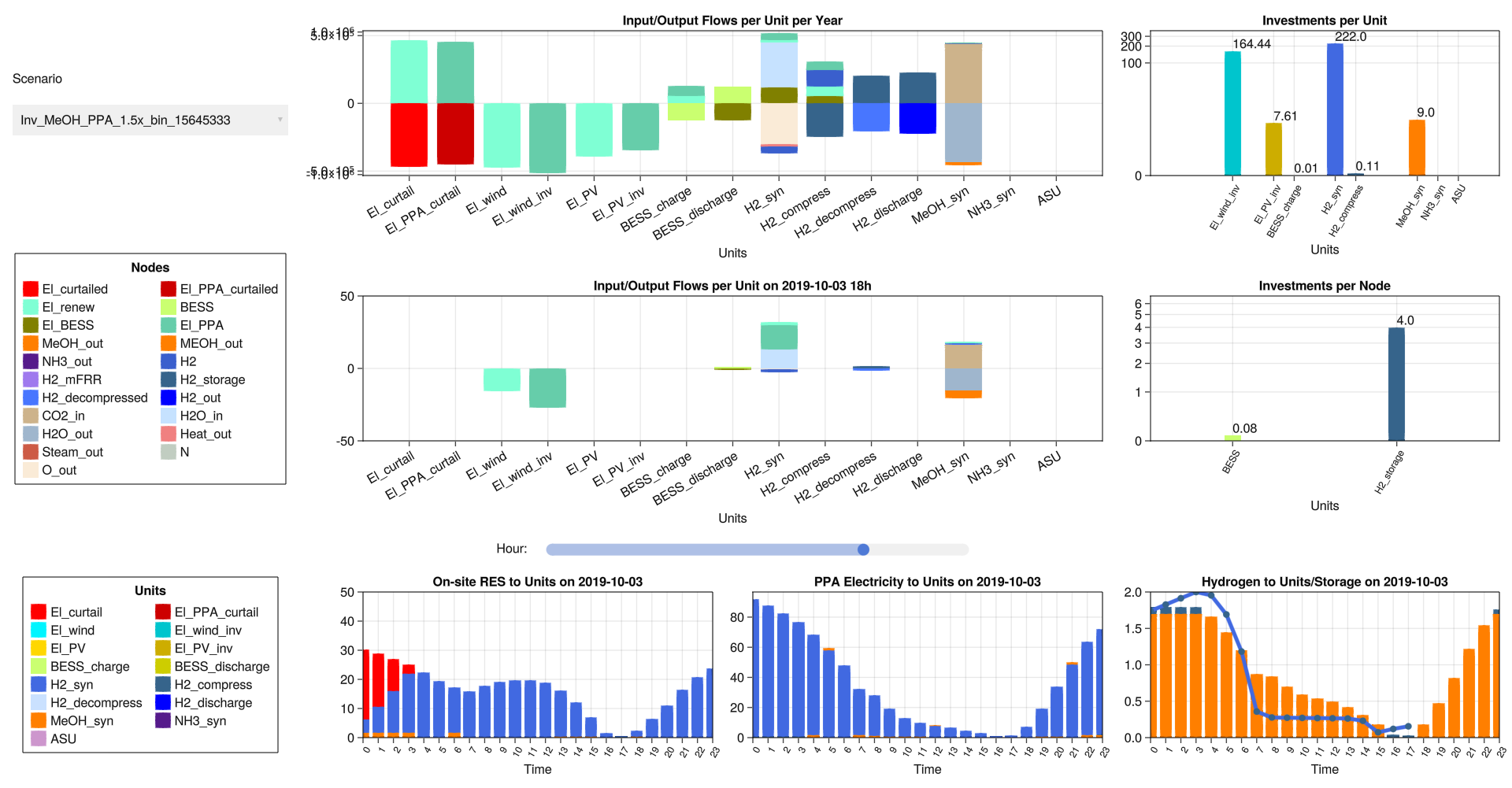} 
\caption{GreenLab Skive energy park investment and illustrative operational decisions of scenario S7 in Table~\ref{tab:scenario_description} (non-indicated values are zero).}
\label{fig:db_s7}
\end{figure*}

\begin{figure*}[ht]
\centering
\includegraphics[width=\textwidth, trim = {15, 15, 15, 15}, clip]{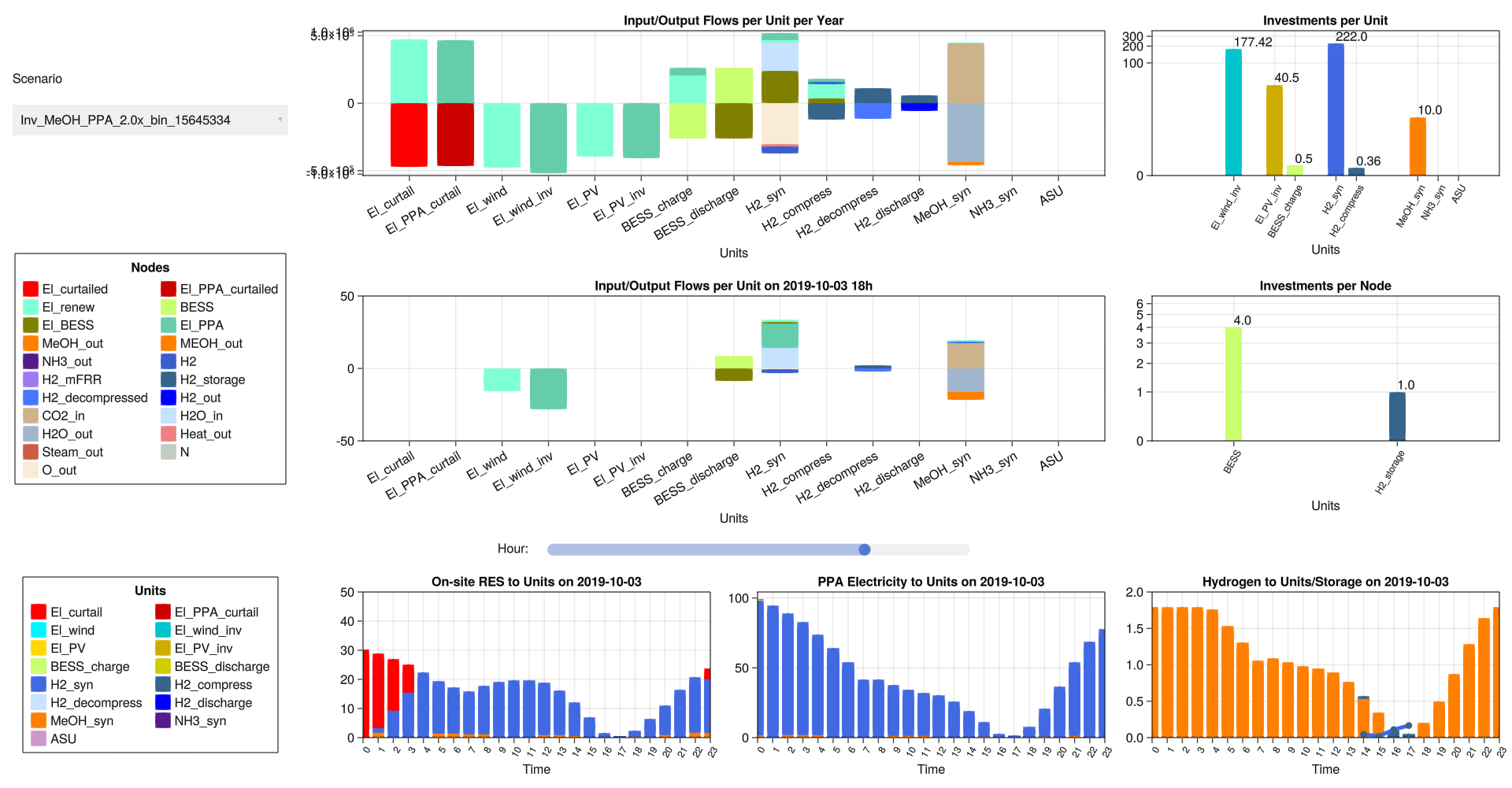} 
\caption{GreenLab Skive energy park investment and illustrative operational decisions of scenario S8 in Table~\ref{tab:scenario_description} (non-indicated values are zero).}
\label{fig:db_s8}
\end{figure*}

\begin{figure*}[ht]
\centering
\includegraphics[width=\textwidth, trim = {15, 15, 15, 15}, clip]{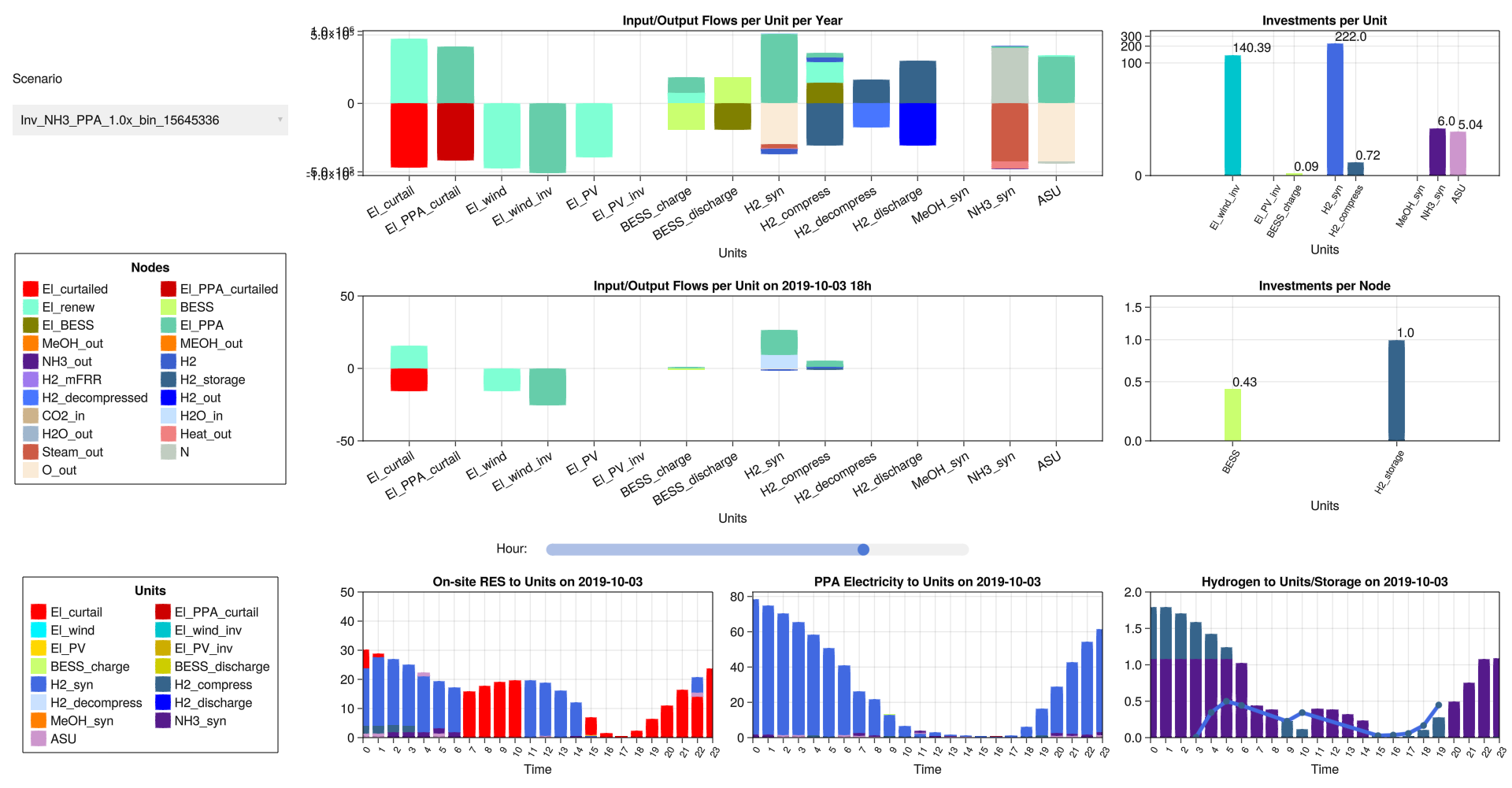} 
\caption{GreenLab Skive energy park investment and illustrative operational decisions of scenario S9 in Table~\ref{tab:scenario_description} (non-indicated values are zero).}
\label{fig:db_s9}
\end{figure*}

\begin{figure*}[ht]
\centering
\includegraphics[width=\textwidth, trim = {15, 15, 15, 15}, clip]{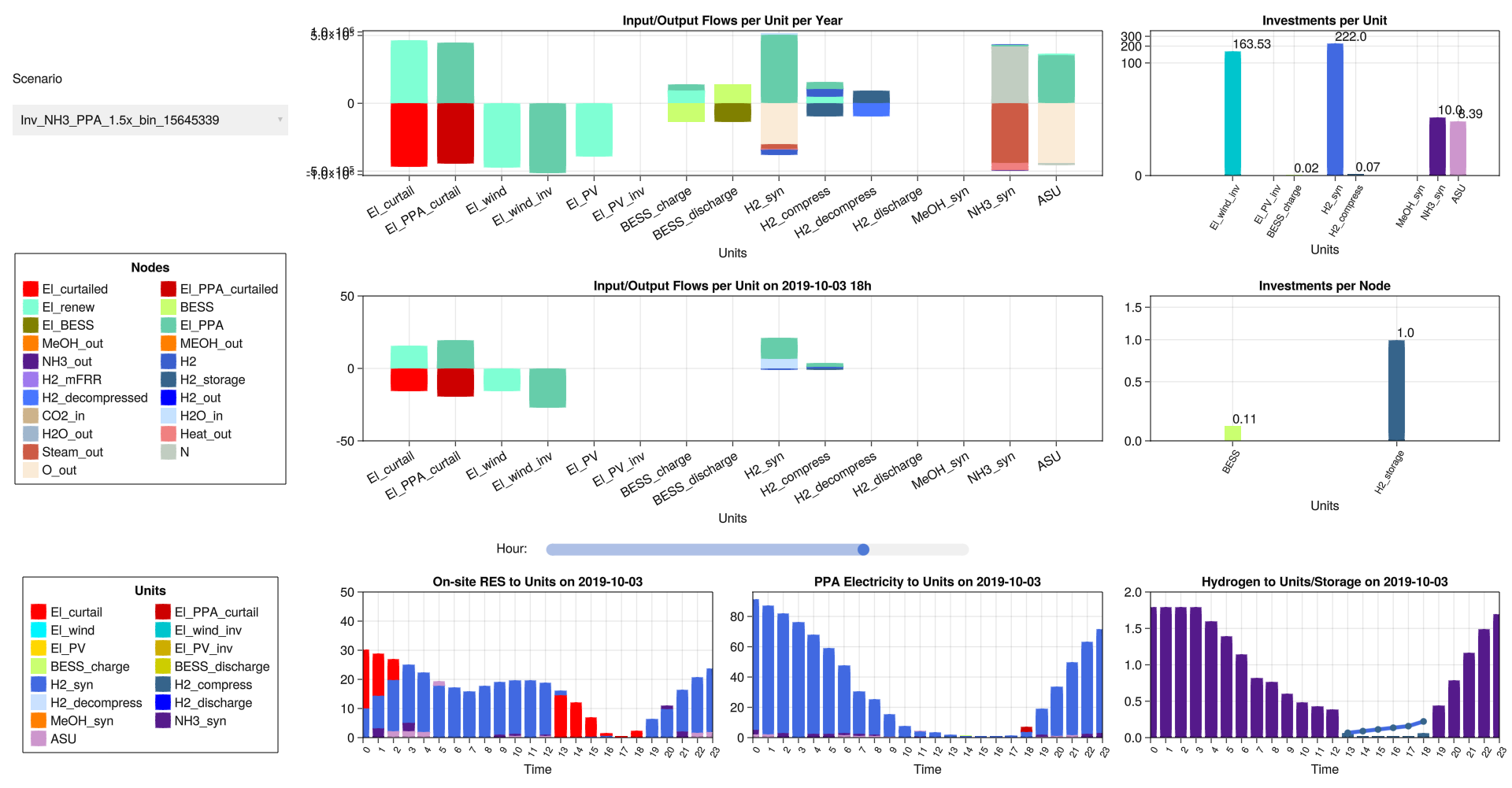} 
\caption{GreenLab Skive energy park investment and illustrative operational decisions of scenario S10 in Table~\ref{tab:scenario_description} (non-indicated values are zero).}
\label{fig:db_s10}
\end{figure*}

\begin{figure*}[ht]
\centering
\includegraphics[width=\textwidth, trim = {15, 15, 15, 15}, clip]{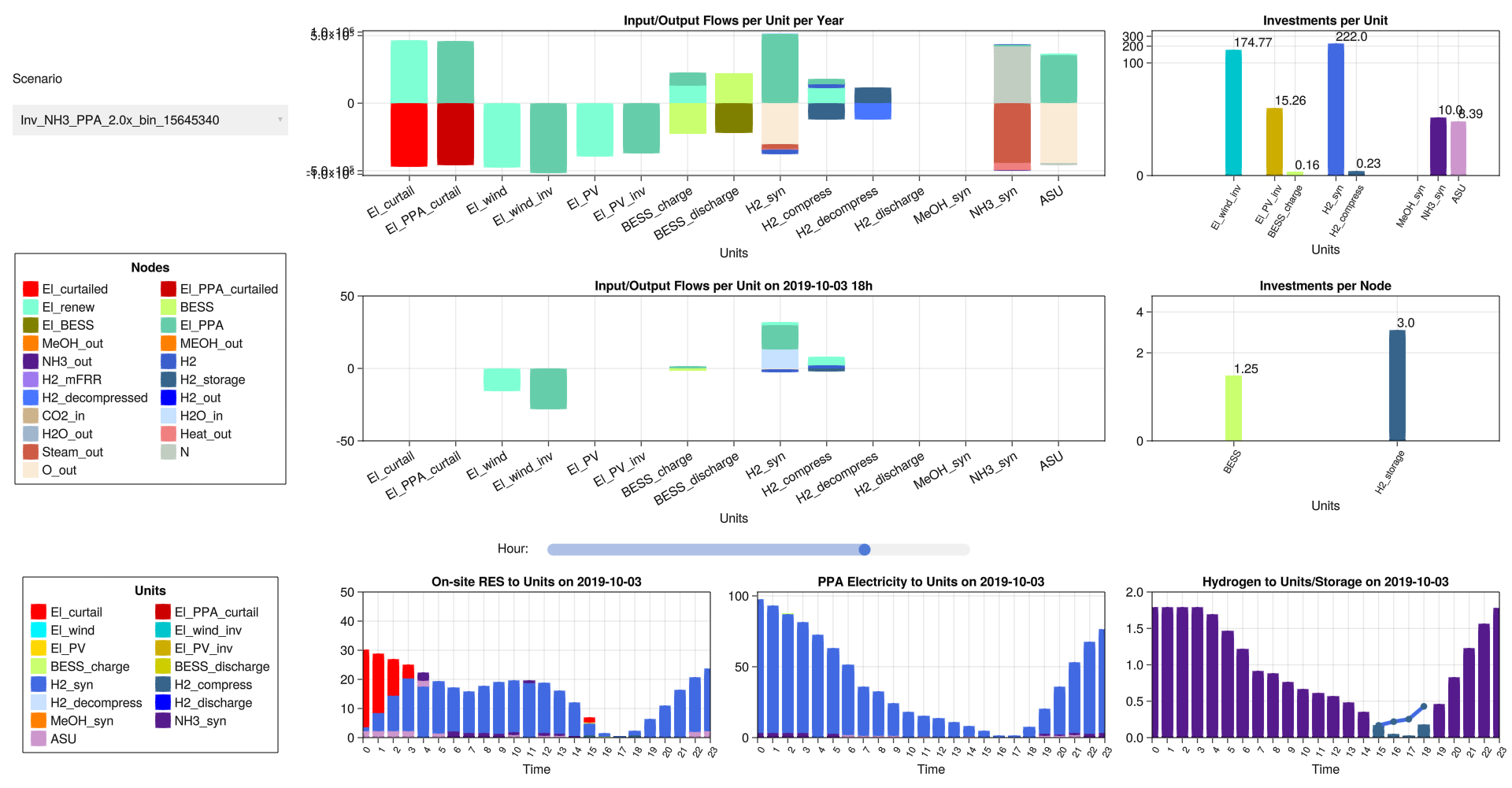} 
\caption{GreenLab Skive energy park investment and illustrative operational decisions of scenario S11 in Table~\ref{tab:scenario_description} (non-indicated values are zero).}
\label{fig:db_s11}
\end{figure*}

We investigate four groups of scenarios (Tab.~\ref{tab:scenario_description}) with results (Tab. \ref{tab:results}) illustrated in Fig.~\ref{fig:investresults} and with more detail in Figs.\ref{fig:db_s6}--\ref{fig:db_s9-cont}.

\begin{table}[h]
    \begin{center} 
        \scalebox{1}{
            \begin{tabular}{ | c | c | c | c |} 
               \hline
               \textbf{\#} & \textbf{Fuels} & \textbf{RES} & \textbf{Fuel Prices} \\ \hline
                0 & H\textsubscript{2} & Local & 1x \\
                1 & H\textsubscript{2} & Local       & 1.5x         \\
                2 & H\textsubscript{2} & Local       & 2x    \\ \hline
                3 & H\textsubscript{2}  & Local + PPA & 1x \\
                4 & H\textsubscript{2}  & Local + PPA & 1.5x     \\
                5 & H\textsubscript{2}  & Local + PPA & 2x  \\ \hline
                6 & H\textsubscript{2} + MeOH  & Local + PPA & 1x \\
                7 & H\textsubscript{2} + MeOH  & Local + PPA & 1.5x     \\
                8 & H\textsubscript{2} + MeOH  & Local + PPA & 2x  \\ \hline
                9 & H\textsubscript{2} + NH\textsubscript{3}  & Local + PPA & 1x \\
                10 & H\textsubscript{2} + NH\textsubscript{3}   & Local + PPA & 1.5x     \\
                11 & H\textsubscript{2} + NH\textsubscript{3}  & Local + PPA & 2x  \\ \hline
            \end{tabular}
        }
    \end{center}
    \caption{Scenario descriptions.}
    \label{tab:scenario_description}
    \vspace{-20pt}
\end{table}

\subsection{Scenario Results}
The first two groups (S0--5), allow only hydrogen production. In S0--2, only the local RES (27\,MW of PV and 54\,MW of wind) are available. In S3--5, investments in additional RES of the same kind can be made. These approximate a PPA considering additionality as well as hourly and regional matching (similar to \cite{zeyen_hourly_2022}). In S6--8, investments in methanol production, in S9-11, in ammonia production instead, are allowed. The base prices are P2X price projections made for 2030 by the Danish Energy Agency: H\textsubscript{2} for \euro 2.16 per kg, MeOH for \euro 0.65 per kg, and NH\textsubscript{3} for \euro 0.46 per kg \cite{energinet_system_2022}. In addition, we examine price increases of 50\% and 100\%---roughly equivalent to the increase that natural-gas-based ammonia reached in 2021 and 2022, respectively.

\paragraph{\textbf{S0--2 (H\textsubscript{2} Local RES)}} Using the base fuel prices and with only the local RES available (S0), the system does not invest in any production capacity. We assume the P2X owner has to pay the 2019 DK1 day-ahead prices (without additional tariffs) for local electricity---then H\textsubscript{2} production is not profitable. With increased H\textsubscript{2} prices (S1+2), the synthesizer and compressor capacities are expanded up to 110 electrolyzer stacks with an overall capacity of 0.89\,t\textsubscript{H\textsubscript{2}} per hour and a compressor of the same size to sell the compressed H\textsubscript{2}.

\paragraph{\textbf{S3--5 (H\textsubscript{2} PPA)}} Allowing contracting PPAs, investments in additional wind generation are made (up to 171\,MW). For all fuel prices, the maximum number of 222~H\textsubscript{2} stacks are invested in (with a capacity of around 1.8\,t\textsubscript{H\textsubscript{2}}  per hour). Some negligible investments in battery capacity are made to fulfil the (small) minimum load requirement of 20\% per single stack.

\paragraph{\textbf{S6--8 (MeOH PPA)}} Allowing investments in MeOH production, similar investments in wind are made (up to 177\,MW). For the high price scenarios (S7+8) up to 40\,MW of additional PV are invested in. Only with +100\% fuel prices (e.g., MeOH \euro 1,300 per ton) (S8), the maximum available capacity of 10\,t\textsubscript{MeOH}  per hour is invested in. Base prices (S6) lead to a capacity of 4\,t\textsubscript{MeOH}  per hour, selling a mix of MeOH (86\%) and compressed H\textsubscript{2} (14\%). Higher fuel prices lead to more MeOH and, eventually, no H\textsubscript{2} being sold.

At base prices (S6), two H\textsubscript{2} storage tanks of 500\,kg\textsubscript{H\textsubscript{2}} each are invested in to buffer H\textsubscript{2} supply to the MeOH synthesizer. A smaller compressor of 1.05\,t\textsubscript{H\textsubscript{2}} per hour is sufficient because most H\textsubscript{2} is directly used for MeOH production. A 50\% increase in fuel prices (S7) leads to four H\textsubscript{2} storage units (2\,t\textsubscript{H2}) which are subsequently filled using a small compressor unit of 0.11\,t\textsubscript{H2}/h. A 100\% increase in fuel prices (S8) reduces the need for H2 storage (500\,kg\textsubscript{H2} only). Instead, the additional PV and a BESS of 4\,MW is used to buffer some low-wind hours to continuously feed H\textsubscript{2} to the MeOH units.

\paragraph{\textbf{S9--11 (NH\textsubscript{3} PPA)}} Allowing for NH\textsubscript{3} production, wind investments are similar (up to 174\,MW). Base prices (S9) lead to investments of 6\,t\textsubscript{NH\textsubscript{3}} per hour with mostly NH\textsubscript{3} being sold (93\%), higher fuel prices lead to the maximum capacity of 10\,t\textsubscript{NH\textsubscript{3}} per hour and all H\textsubscript{2} being used for NH\textsubscript{3} production.
Hydrogen supply is mainly buffered using H\textsubscript{2} storage (500\,kg for S10 and 1.5\,t\textsubscript{H\textsubscript{2}} for S11) plus some PV and BESS in S11.

\subsection{Sensitivities} \label{sec:sensitivity}
In addition, we investigate how the given results change when the CAPEX of the MeOH and NH\textsubscript{3} units are lower than expected, and when no unit commitment is enforced. 

\begin{figure*}[ht]
\centering
\includegraphics[width=\textwidth, trim = {15, 15, 15, 15}, clip]{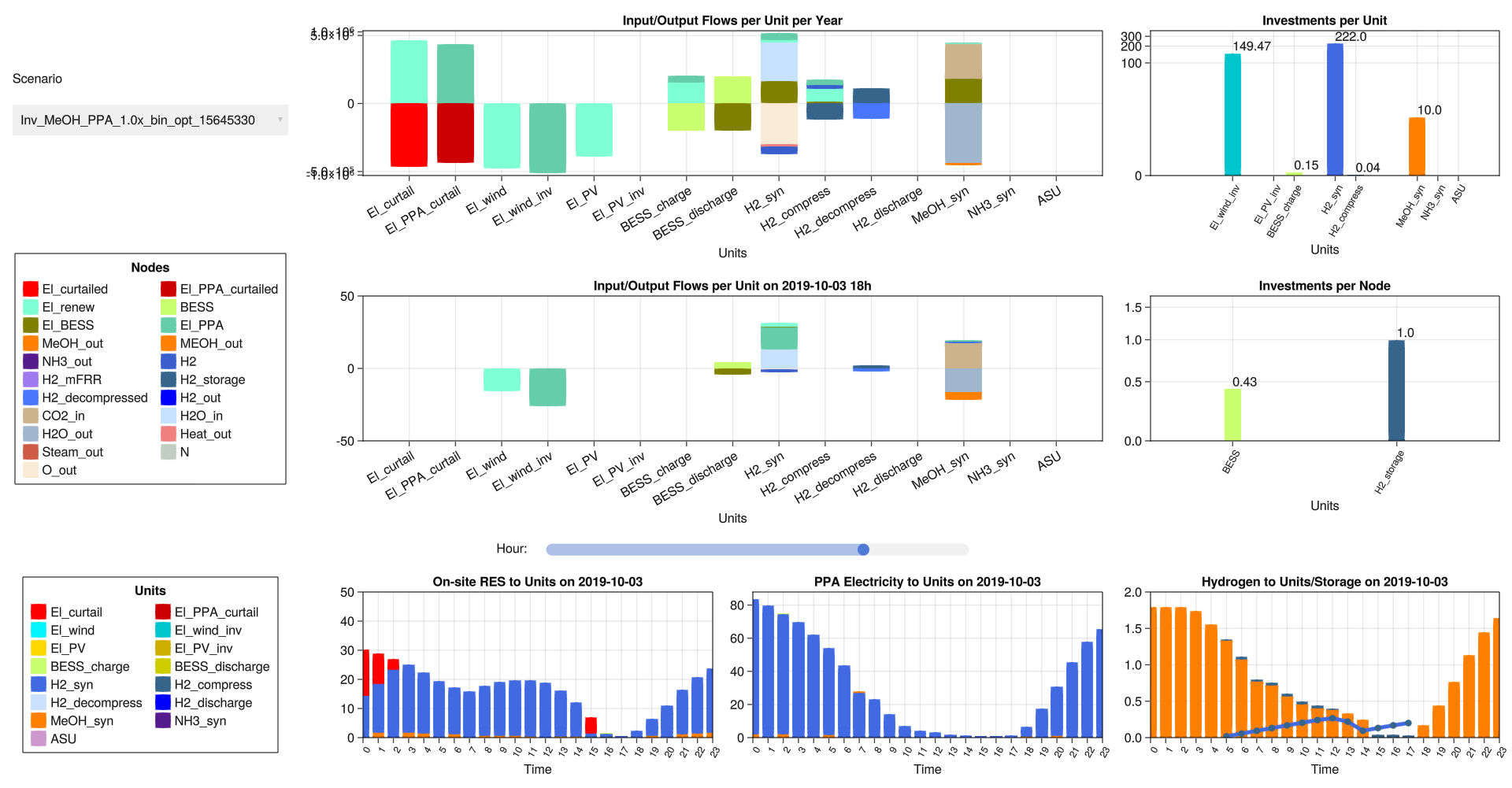} 
\caption{GreenLab Skive energy park investment and illustrative operational decisions of scenario S6 \textbf{with optimistic CAPEX} (non-indicated values are zero).}
\label{fig:db_s6-opt}
\end{figure*}

\begin{figure*}[ht]
\centering
\includegraphics[width=\textwidth, trim = {15, 15, 15, 15}, clip]{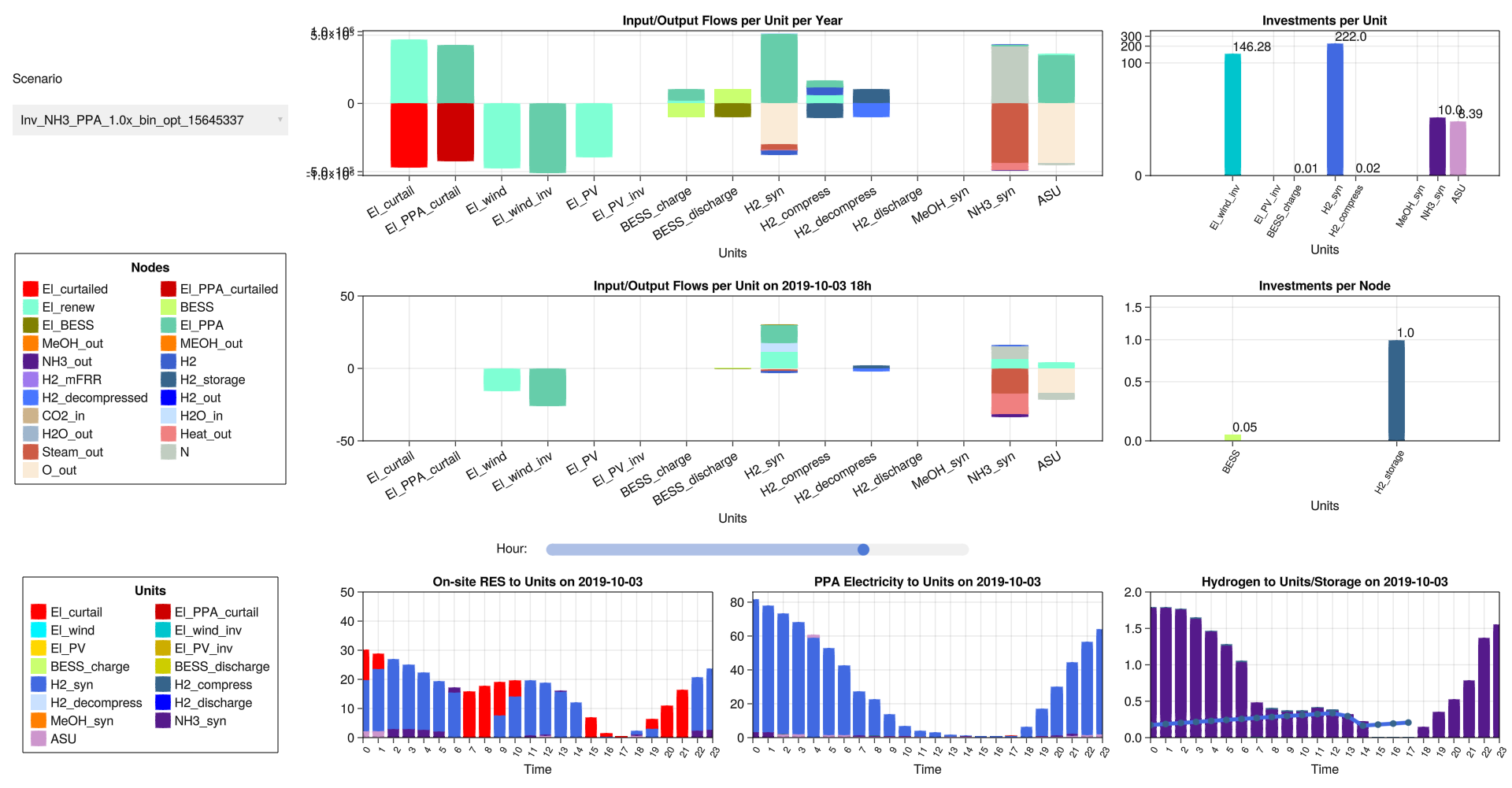} 
\caption{GreenLab Skive energy park investment and illustrative operational decisions of scenario S9 \textbf{with optimistic CAPEX} (non-indicated values are zero).}
\label{fig:db_s9-opt}
\end{figure*}

\paragraph{\textbf{Optimistic P2X CAPEX}}
The values used so far are based on Danish Energy Agency estimates for a 12.5\,t per hour plant size \cite{danish_energy_agency_technology_2022}.
In general, investment expenditures quoted vary greatly between 4--17\,M\euro~per ton methanol \cite{pypsa_technology_2022} and 3--7\,M\euro~per ton ammonia \cite{ikaheimo_power--ammonia_2018} per hour.
We find that with more optimistic CAPEX, full investment in P2X assets (even with base fuel prices) is conducted (S9--11). PPAs of 150\,MW and less H\textsubscript{2} storage assets are needed to secure H\textsubscript{2} supply. Almost only P2X fuels and no H\textsubscript{2} is sold (Figs. \ref{fig:db_s6-opt}+\ref{fig:db_s9-opt}).

\begin{figure*}[ht]
\centering
\includegraphics[width=\textwidth, trim = {15, 15, 15, 15}, clip]{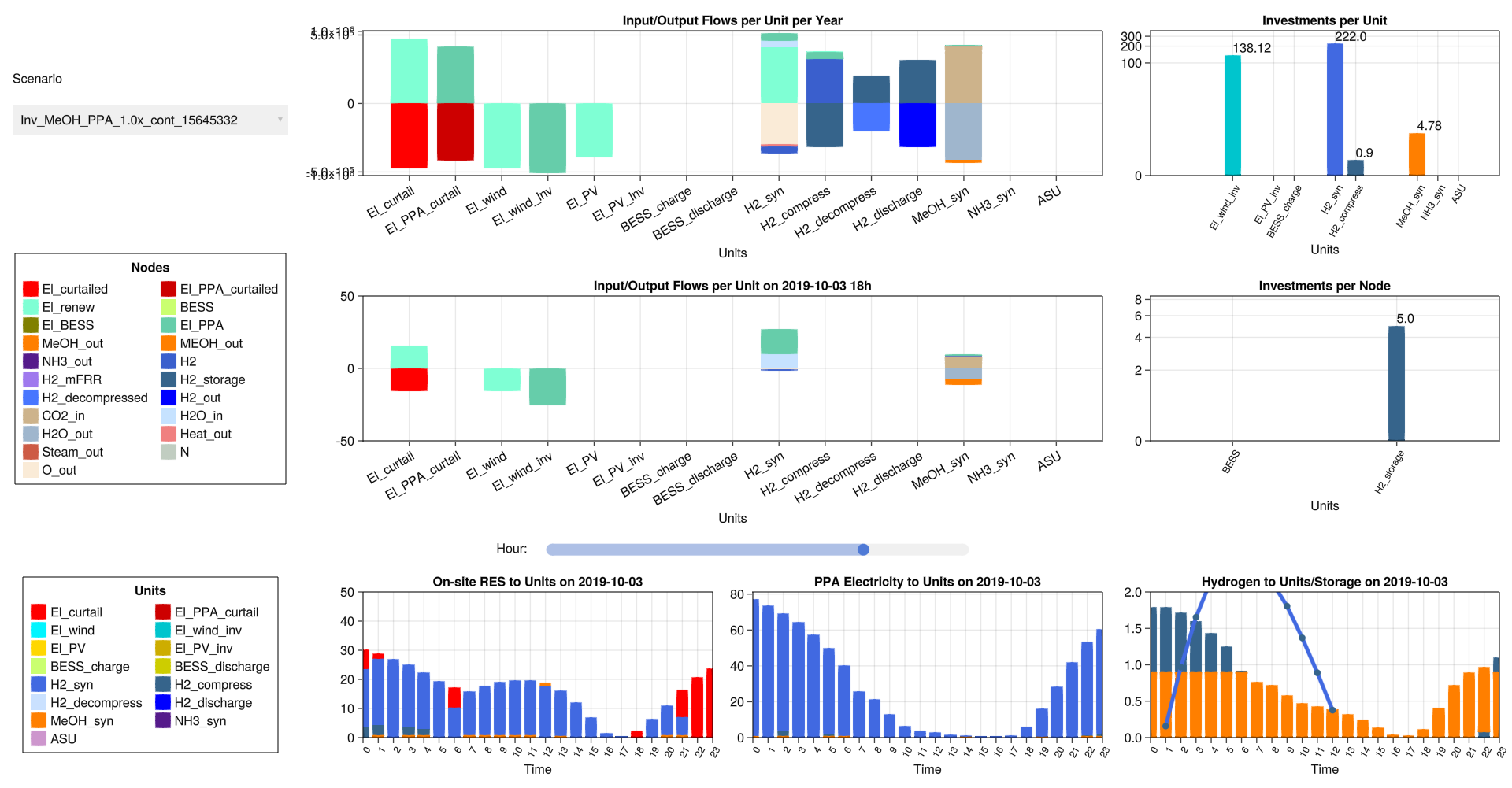} 
\caption{GreenLab Skive energy park investment and illustrative operational decisions of scenario S6 \textbf{without unit commitment }(non-indicated values are zero).}
\label{fig:db_s6-cont}
\end{figure*}

\begin{figure*}[ht]
\centering
\includegraphics[width=\textwidth, trim = {15, 15, 15, 15}, clip]{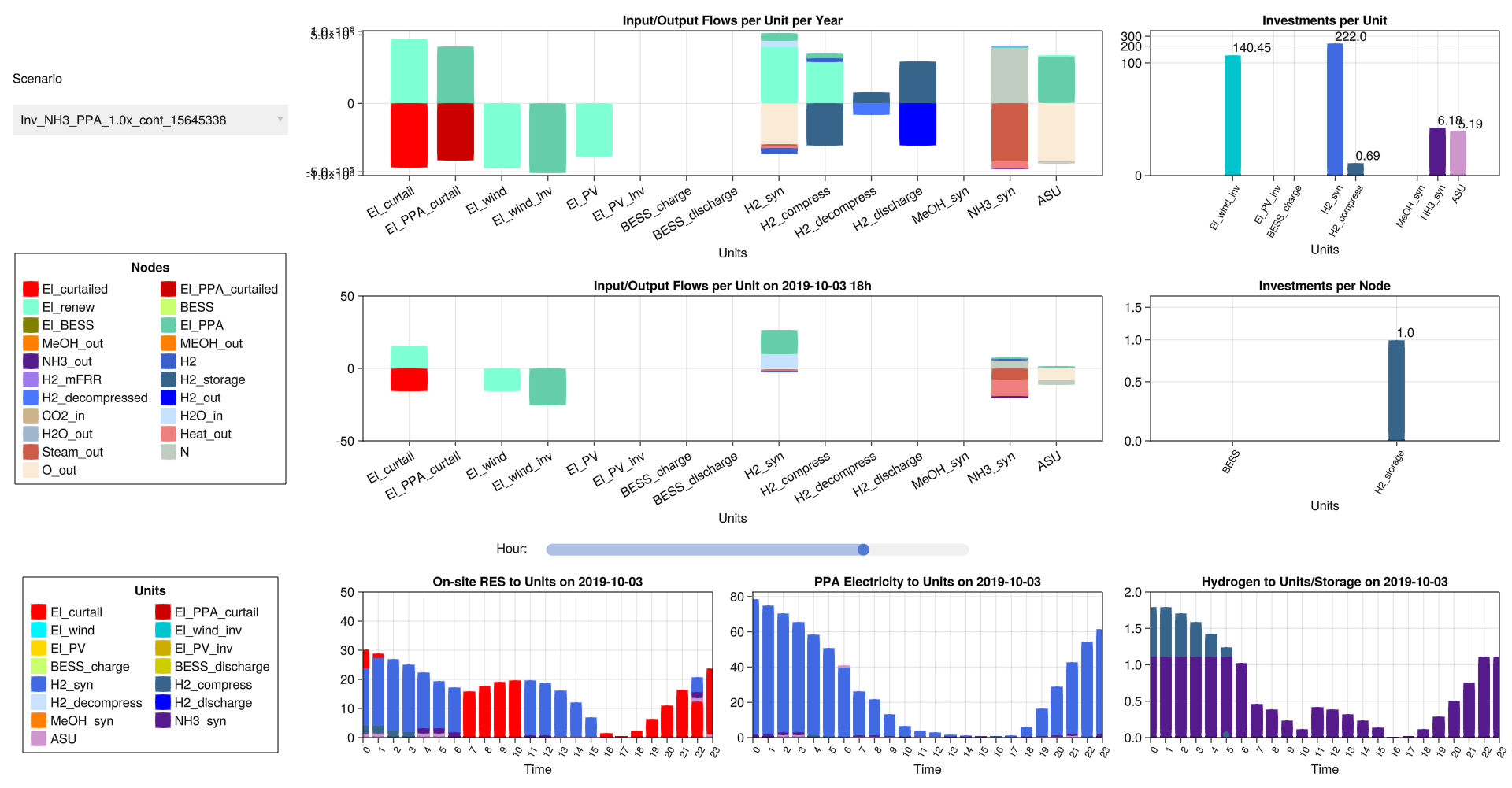} 
\caption{GreenLab Skive energy park investment and illustrative operational decisions of scenario S9 \textbf{without unit commitment }(non-indicated values are zero).}
\label{fig:db_s9-cont}
\end{figure*}

\paragraph{\textbf{No Unit Commitment}}
For the H\textsubscript{2}-only scenarios (S0--5), the linear optimization delivers approximately the same results as the MILP version because unit commitment plays a minor role for the more flexible electrolyzer. For the P2X scenarios (S6--11), however, the need for storage flexibility varies showing complex trade-offs between excess capacities, H\textsubscript{2} storage, and BESS (Figs. \ref{fig:db_s6-cont}--\ref{fig:db_s9-cont}). The underlying effects should be investigated further.

\begin{table}
    \begin{center} 
        \scalebox{1}{
            \begin{tabular}{ | c | c | c | c | c |} 
               \hline
               \textbf{\#} & \textbf{H2 sold} & \textbf{H2 produced}& \textbf{P2X sold} & \textbf{Profit} \\ 
                & t & t (FLH)  & t (FLH) & M\euro \\ \hline
                0 & 0 & 0 & - & - \\
                1 & all & 4,532t (5,709h)   &-   & 2.73        \\
                2 & all & 4,948t (5,553h) & -       & 7.94    \\ \hline
                3 & all & 11,509t (6400h)  & - & 4.96 \\
                4 &  all & 13,134t (7,304h) & - & 18.45    \\
                5 & all & 13,457t (7,483h)  & - & 32.86  \\ \hline
                6 & 5,541t & 11,846t (6,587h) & 33,503t (8,376h)  & \textbf{5.68} \\
                7 & 390t  & 13,570t (7,546h) & 69,959t (7,773h) & \textbf{27.56 }    \\
                8 & 3t & 13,968t (7,768h) & 74,281t (7,428h) & \textbf{50.67} \\ \hline
                9 & 3,434t & 11,905t (6,620h) & 47,038t (7,840h) & 5.66 \\
                10 & 0  & 13,357t (7,428h) & 74,202t (7,420h) & 21.86    \\
                11 & 0 & 13,677t (7,606h) &  75,981t (7,598h) & 39.02 \\ \hline
            \end{tabular}
        }
    \end{center}
    \caption{Produced and sold fuel, full-load hours and profits.}
    \label{tab:results}
    \vspace{-20pt}
\end{table}

\section{Conclusion}
We model the capacity expansion problem for an energy hub inspired by the industrial park GreenLab Skive in North-West Denmark. Following the \gls{EC}'s regulation on renewable hydrogen, we consider additionality as well as hourly and regional matching. The site has (fixed capacity) local RES (which it purchases at day-ahead prices) and can invest in additional PPAs and production capacities for different P2X assets. We examine scenarios with only H\textsubscript{2}, and added either MeOH or NH\textsubscript{3} production. In addition, we evaluate the need for flexibility considering H\textsubscript{2} and battery storage systems. We take into account current P2X fuel price projections for 2030 and potential increases by 50\% and 100\%.

Using the current P2X fuel price projections for 2030 and restricting electricity sourcing to the existing local RES, only producing H\textsubscript{2} is not profitable (at \euro 2.16 per kg). Only with higher fuel prices, a PPA, or producing H\textsubscript{2} derivatives can a profitable business case be achieved. 

Producing P2X fuels, the PPA volume remains similar to the H\textsubscript{2}-only case because electricity is mainly used to produce and compress H\textsubscript{2}. 50\% higher fuel prices lead to a mix of additional PV and H\textsubscript{2} storage to buffer supply to the P2X units. With higher fuel prices (+100\%), continuous H\textsubscript{2} supply is secured using additional PV and battery storage. In all price scenarios, producing MeOH leads to the highest profit, followed by NH\textsubscript{3} and H\textsubscript{2}---the impact of CO\textsubscript{2} prices and availability should be investigated. More optimistic CAPEX shows larger assets already at lower fuel prices, reducing the need for flexibility from H\textsubscript{2} or BESS.

The maximum available capacity of the electrolyzer drives the PPA volume, which is needed to comply with renewable fuel regulations instead of sourcing from the grid. With additional RES available via PPA, the maximum electrolyzer installation seems to be the no-regret option (with quite high full-load hours). When producing the more profitable P2X fuels, H\textsubscript{2} supply can be subsequently expanded starting with a small H\textsubscript{2} storage unit of 500 kg\textsubscript{H\textsubscript{2}} plus any required fuel storage depending on the final demand pattern, complemented later with additional PV and BESS.

Because we do not allow selling excess electricity, substantial amounts of RES are "curtailed". This could provide additional income for the owner of the assets---leading to an upper-bound price assumption for the PPA. However, large amounts of the local RES are also not consumed during hours of high day-ahead prices. This indicates that the balance of risks and rewards of these bilateral contracts needs to be reassessed in the context of instruments with greater long-term price certainty, such as PPAs on the supply side, but also in the light of the upcoming European Hydrogen Bank auctions for carbon contracts for difference (CCFDs) on the demand side. The impact on overall system capacities, flexibility, and emissions should be examined in this context.

 \printglossary[type=\acronymtype]

 \bibliographystyle{vancouver} 
 \bibliography{references}

\begin{thebibliography}{10}

\bibitem{european_commission_commission_2023-1}
{European Commission}. Commission sets out rules for renewable hydrogen; 2023.
\newblock Available from:
  \url{https://ec.europa.eu/commission/presscorner/detail/en/IP_23_594}.

\bibitem{european_commission_repowereu_2022}
{European Commission}. {REPowerEU}: {A} plan to rapidly reduce dependence on
  {Russian} fossil fuels and fast forward the green transition; 2022.
\newblock Available from:
  \url{https://ec.europa.eu/commission/presscorner/detail/en/IP_22_3131}.

\bibitem{european_commission_commission_2023}
{European Commission}. Commission outlines {European} {Hydrogen} {Bank} to
  boost renewable hydrogen; 2023.
\newblock Available from: \url{https://europa.eu/!kBjCrb}.

\bibitem{european_commission_european_2021}
{European Commission}. European {Green} {Deal}: {Commission} proposes
  transformation of {EU} economy and society to meet climate ambitions; 2021.
\newblock Available from:
  \url{https://ec.europa.eu/commission/presscorner/detail/en/ip_21_3541}.

\bibitem{european_commission_directive_2018}
{European Commission}. Directive ({EU}) 2018/2001 of the {European}
  {Parliament} and of the {Council} of 11 {December} 2018 on the promotion of
  the use of energy from renewable sources; 2018.
\newblock Available from:
  \url{https://eur-lex.europa.eu/legal-content/EN/TXT/?uri=uriserv:OJ.L_.2018.328.01.0082.01.ENG&toc=OJ:L:2018:328:TOC}.

\bibitem{world_resources_institute_wri_ghg_2023}
{World Resources Institute (WRI)}. {GHG} {Protocol} {Standards} {Update}
  {Process}: {Topline} {Findings} {From} {Scope} 2 {Feedback}; 2023.
\newblock Available from:
  \url{https://ghgprotocol.org/sites/default/files/2023-05/Topline%20Findings%20from%20Scope%202%20Feedback%20Webinar_GHG%20Protocol_05.02.2023.pdf}.

\bibitem{european_commission_green_2023}
{European Commission}. The {Green} {Deal} {Industrial} {Plan}: putting
  {Europe}'s net-zero industry in the lead; 2023.
\newblock Available from:
  \url{https://ec.europa.eu/commission/presscorner/detail/en/ip_23_510}.

\bibitem{european_commission_questions_2023}
{European Commission}. Questions and {Answers} on the {EU} {Delegated} {Acts}
  on {Renewable} {Hydrogen}; 2023.
\newblock Available from:
  \url{https://ec.europa.eu/commission/presscorner/detail/en/qanda_23_595}.

\bibitem{ricks_minimizing_2023}
Ricks W, Xu Q, Jenkins JD.
\newblock Minimizing emissions from grid-based hydrogen production in the
  {United} {States}.
\newblock Environmental Research Letters. 2023 Jan;18(1):014025.
\newblock Publisher: IOP Publishing.
\newblock Available from: \url{https://dx.doi.org/10.1088/1748-9326/acacb5}.

\bibitem{brauer_green_2022}
Brauer J, Villavicencio M, Trüby J.
\newblock Green hydrogen – {How} grey can it be?
\newblock Robert Schuman Centre for Advanced Studies (FSR); 2022.
\newblock Available from: \url{https://cadmus.eui.eu/handle/1814/74850}.

\bibitem{zeyen_hourly_2022}
Zeyen E, Riepin I, Brown T. Hourly versus annually matched renewable supply for
  electrolytic hydrogen; 2022.
\newblock Available from:
  \url{https://zenodo.org/record/7457441/preview/Report_TUB_hourlyvsannually.pdf}.

\bibitem{kountouris_power--x_2023}
Kountouris I, Langer L, Bramstoft R, Münster M, Keles D.
\newblock Power-to-{X} in energy hubs: {A} {Danish} case study of renewable
  fuel production.
\newblock Energy Policy. 2023 Apr;175:113439.
\newblock Available from:
  \url{https://www.sciencedirect.com/science/article/pii/S0301421523000241}.

\bibitem{ihlemann_spineopt_2022}
Ihlemann M, Kouveliotis-Lysikatos I, Huang J, Dillon J, O’Dwyer C, Rasku T,
  et~al.
\newblock {SpineOpt}: {A} flexible open-source energy system modelling
  framework.
\newblock Energy Strategy Reviews. 2022 Sep;43:100902.
\newblock Available from:
  \url{https://www.sciencedirect.com/science/article/pii/S2211467X22000955}.

\bibitem{kiviluoma_spine_2022}
Kiviluoma J, Pallonetto F, Marin M, Savolainen PT, Soininen A, Vennström P,
  et~al.
\newblock Spine {Toolbox}: {A} flexible open-source workflow management system
  with scenario and data management.
\newblock SoftwareX. 2022 Jan;17:100967.
\newblock Available from:
  \url{https://www.sciencedirect.com/science/article/pii/S2352711021001886}.

\bibitem{energinet_system_2022}
{Energinet}. {SYSTEM} {PERSPECTIVE} {ANALYSIS} 2022 {PATHWAYS} {TOWARDS} {A}
  {ROBUST} {FUTURE} {ENERGY} {SYSTEM}; 2022.
\newblock Available from:
  \url{https://www.energinet.dk/media/y5rhoqjy/pathways-towards-a-robust-future-energy-system_energinet-2023-01-23.pdf}.

\bibitem{danish_energy_agency_technology_2022}
{Danish Energy Agency}. Technology {Data} for energy carrier generation and
  conversion; 2022.
\newblock Available from:
  \url{https://ens.dk/en/our-services/projections-and-models/technology-data/technology-data-renewable-fuels}.

\bibitem{pypsa_technology_2022}
{PyPSA}. Technology data 2030; 2022.
\newblock Available from:
  \url{https://github.com/PyPSA/technology-data/blob/master/outputs/costs_2030.csv}.

\bibitem{ikaheimo_power--ammonia_2018}
Ikäheimo J, Kiviluoma J, Weiss R, Holttinen H.
\newblock Power-to-ammonia in future {North} {European} 100 \% renewable power
  and heat system.
\newblock International Journal of Hydrogen Energy. 2018 Sep;43(36):17295-308.
\newblock Available from:
  \url{https://linkinghub.elsevier.com/retrieve/pii/S0360319918319931}.

\end{thebibliography}

\clearpage
\appendices

\section{Additional Figures}

\FloatBarrier
\section{Data Input} \label{sec:data-input}

\begin{table}[ht]
    \begin{center} 
        \scalebox{1}{
            \begin{tabular}{ | c | c | c | c | c | c | c | c | c | c | c | c | c | c | c |} 
               \hline
               Scenario & 0 & 1 & 2 & 3 & 4 & 5 & 6 & 7 & 8 & 9 & 10 & 11 & a & b  \\ \hline
Base                                    & \checkmark & \checkmark  &\checkmark & \checkmark & \checkmark  &\checkmark & \checkmark & \checkmark  &\checkmark  & \checkmark & \checkmark  &\checkmark & \checkmark  &\checkmark  \\
Inv                                     &  \checkmark &\checkmark   & \checkmark & \checkmark & \checkmark  &\checkmark  & \checkmark & \checkmark  &\checkmark & \checkmark & \checkmark  &\checkmark& \checkmark  &\checkmark   \\
Tech-H2-units (place before H2 invests) & \checkmark  &\checkmark   & \checkmark & \checkmark & \checkmark  &\checkmark & \checkmark & \checkmark  &\checkmark & \checkmark & \checkmark  &\checkmark & \checkmark  &\checkmark  \\
Inv-H2-on                               &  \checkmark & \checkmark  & \checkmark & \checkmark & \checkmark  &\checkmark & \checkmark & \checkmark  &\checkmark & \checkmark & \checkmark  &\checkmark & \checkmark  &\checkmark  \\
Inv-H2-int                              &  \checkmark &\checkmark   & \checkmark & \checkmark & \checkmark  &\checkmark & \checkmark & \checkmark  &\checkmark & \checkmark & \checkmark  &\checkmark &&   \\
Inv-H2-cont                              &&&&&&&&&&&& & \checkmark  &\checkmark   \\
Inv-MeOH-on                             &  &  & & &  & & \checkmark & \checkmark  &\checkmark  & & && (\checkmark) & (\checkmark)\\
Inv-MeOH-bin (combine with on)          &  &  & & &  & & \checkmark & \checkmark  &\checkmark & & & &&\\
Inv-MeOH-cont (combine with on)      &&&    &  &  & & &  & & & & & (\checkmark) & (\checkmark)  \\
Inv-NH3-on                             &  &  & & & & & & & & \checkmark & \checkmark  &\checkmark & (\checkmark) & (\checkmark)  \\
Inv-NH3-bin (combine with on)          &  &  & & & & & & & & \checkmark & \checkmark  &\checkmark && \\
Inv-NH3-cont (combine with on)            &  & & & & & & & & &&& & (\checkmark) & (\checkmark)  \\
Inv-PPA                                 &  &  & & \checkmark & \checkmark  &\checkmark & \checkmark & \checkmark  &\checkmark & \checkmark & \checkmark  &\checkmark  & (\checkmark) & (\checkmark)  \\
Inv-storage-compress                    &  \checkmark &\checkmark   & \checkmark & \checkmark & \checkmark  &\checkmark & \checkmark & \checkmark  &\checkmark & \checkmark & \checkmark  &\checkmark  & \checkmark  &\checkmark  \\
Mod-UC                                  & \checkmark  & \checkmark  & \checkmark & \checkmark & \checkmark  &\checkmark & \checkmark & \checkmark  &\checkmark  & \checkmark & \checkmark  &\checkmark & \checkmark  & \\
Premium-1.5x                            &  & \checkmark  & & &\checkmark & & & \checkmark& & &\checkmark && (\checkmark) & (\checkmark)\\
Premium-2x                              &  &  & \checkmark & & & \checkmark& & &\checkmark & & &\checkmark& (\checkmark) & (\checkmark)\\
Solver-Gurobi                           &  \checkmark & \checkmark  & \checkmark & \checkmark & \checkmark  &\checkmark & \checkmark & \checkmark  &\checkmark & \checkmark & \checkmark  &\checkmark& \checkmark  &\checkmark \\
Year                                    &  \checkmark & \checkmark  & \checkmark & \checkmark & \checkmark  &\checkmark & \checkmark & \checkmark  &\checkmark & \checkmark & \checkmark  &\checkmark& \checkmark  &\checkmark \\ \hline
            \end{tabular}
        }
    \end{center}
    \caption{Paper scenarios matched with model scenarios (stacked).}
    \label{tab:scenario_matching}
\end{table}

\begin{table*}[ht]
\begin{center} 
\scalebox{0.9}{
\begin{tabular}{|l|l|l|l|}
\hline
\textbf{Parameter} & \textbf{Type}   & \textbf{Value}      & \textbf{Scenario} \\ \hline
db\_lp\_solver    & model            & Gurobi.jl           & Solver-Gurobi                  \\
db\_mip\_solver   & model            & Gurobi.jl           & Solver-Gurobi                  \\
duration\_unit    & model            & hour                & Base                           \\
model\_end        & model            & 2019-01-01T03:00:00 & Base                           \\
model\_end        & model            & 2019-12-31T00:00:00 & Year                           \\
model\_start      & model            & 2019-01-01T00:00:00 & Base                           \\
resolution        & temporal\_block  & 1h                  & Base                           \\
block\_end        & temporal\_block  & 364D                & Inv                            \\
block\_start      & temporal\_block  & 0h                  & Inv                            \\
resolution        & temporal\_block  & 364D                & Inv                            \\
constraint\_sense & user\_constraint & \textless{}=        & Inv-MeOH-bin (combine with on) \\
right\_hand\_side & user\_constraint & 8736                & Inv-MeOH-bin (combine with on) \\
constraint\_sense & user\_constraint & \textless{}=        & Inv-NH3-bin (combine with on)  \\
right\_hand\_side & user\_constraint & 8736                & Inv-NH3-bin (combine with on)   \\ \hline      
\end{tabular}
    }
\end{center}
\caption{Model parameters.}
\label{tab:model_param}
\end{table*}

\begin{table*}[h]
\begin{center} 
\scalebox{0.9}{
\begin{tabular}{|l|l|l|l|l|}
\hline
\textbf{Unit} & \textbf{Parameter}               & \textbf{Time} & \textbf{Value}                               & \textbf{Scenario}    \\ \hline
BESS\_charge & candidate\_units                 &        & 3                                            & Inv-storage-compress                    \\
BESS\_charge & fom\_cost                        &        & 4800                                         & Inv-storage-compress                    \\
BESS\_charge & number\_of\_units                &        & 0                                            & Inv-storage-compress                    \\
BESS\_charge & unit\_investment\_cost           &        & 23845                                        & Inv-storage-compress                    \\
BESS\_charge & unit\_investment\_variable\_type &        & unit\_investment\_variable\_type\_continuous & Inv-storage-compress                    \\
El\_PV       & fix\_units\_on                   & hourly & time series                                  & Base                                    \\
El\_wind     & fix\_units\_on                   & hourly & time series                                  & Base                                    \\
H2\_compress & candidate\_units                 &        & 35                                           & Inv-storage-compress                    \\
H2\_compress & fom\_cost                        &        & 8339                                         & Inv-storage-compress                    \\
H2\_compress & number\_of\_units                &        & 0                                            & Inv-storage-compress                    \\
H2\_compress & unit\_investment\_cost           &        & 807792                                       & Inv-storage-compress                    \\
H2\_compress & unit\_investment\_variable\_type &        & unit\_investment\_variable\_type\_continuous & Inv-storage-compress                    \\
H2\_syn      & candidate\_units                 &        & 222                                          & Inv-H2-on                               \\
H2\_syn      & fom\_cost                        &        & 3732                                         & Inv-H2-on                               \\
H2\_syn      & min\_down\_time                  &        & 2h                                           & Mod-UC                                  \\
H2\_syn      & number\_of\_units                &        & 0                                            & Inv-H2-on                               \\
H2\_syn      & number\_of\_units                &        & 222                                          & Tech-H2-units (place before H2 invests) \\
H2\_syn      & online\_variable\_type           &        & unit\_online\_variable\_type\_integer        & Mod-UC                                  \\
H2\_syn      & unit\_investment\_cost           &        & 16574                                        & Inv-H2-on                               \\
H2\_syn      & unit\_investment\_variable\_type &        & unit\_investment\_variable\_type\_continuous & Inv-H2-cont                             \\
H2\_syn      & unit\_investment\_variable\_type &        & unit\_investment\_variable\_type\_integer    & Inv-H2-int         \\ 
H2\_syn	& unit\_investment\_variable\_type	&	& unit\_investment\_variable\_type\_continuous & Inv-H2-cont \\
MeOH\_syn     & candidate\_units                 &  & 1                                              & Inv-MeOH-bin (combine with on)  \\
MeOH\_syn     & candidate\_units                 &  & 10                                             & Inv-MeOH-cont (combine with on) \\
MeOH\_syn     & fom\_cost                        &  & 292972                                         & Inv-MeOH-on                     \\
MeOH\_syn     & is\_active                       &  & true                                           & Inv-MeOH-on                     \\
MeOH\_syn     & number\_of\_units                &  & 0                                              & Inv-MeOH-on                     \\
MeOH\_syn     & unit\_investment\_cost           &  & 1477965                                        & Inv-MeOH-on                     \\
MeOH\_syn     & unit\_investment\_variable\_type &  & unit\_investment\_variable\_type\_integer      & Inv-MeOH-bin (combine with on)  \\
MeOH\_syn     & unit\_investment\_variable\_type &  & unit\_investment\_variable\_type\_continuous   & Inv-MeOH-cont (combine with on) \\
MeOH\_syn\_02 & candidate\_units                 &  & 1                                              & Inv-MeOH-bin (combine with on)  \\
MeOH\_syn\_02 & fom\_cost                        &  & 585944                                         & Inv-MeOH-bin (combine with on)  \\
MeOH\_syn\_02 & is\_active                       &  & true                                           & Inv-MeOH-bin (combine with on)  \\
MeOH\_syn\_02 & number\_of\_units                &  & 0                                              & Inv-MeOH-bin (combine with on)  \\
MeOH\_syn\_02 & unit\_investment\_cost           &  & 2955930                                        & Inv-MeOH-bin (combine with on)  \\
MeOH\_syn\_02 & unit\_investment\_variable\_type &  & unit\_investment\_variable\_type\_integer      & Inv-MeOH-bin (combine with on)  \\
... & & & \\
MeOH\_syn\_10 & candidate\_units                 &  & 1                                              & Inv-MeOH-bin (combine with on)  \\
MeOH\_syn\_10 & fom\_cost                        &  & 2929722                                        & Inv-MeOH-bin (combine with on)  \\
MeOH\_syn\_10 & is\_active                       &  & true                                           & Inv-MeOH-bin (combine with on)  \\
MeOH\_syn\_10 & number\_of\_units                &  & 0                                              & Inv-MeOH-bin (combine with on)  \\
MeOH\_syn\_10 & unit\_investment\_cost           &  & 14779652                                       & Inv-MeOH-bin (combine with on)  \\
MeOH\_syn\_10 & unit\_investment\_variable\_type &  & unit\_investment\_variable\_type\_integer      & Inv-MeOH-bin (combine with on)  \\
NH3\_syn      & candidate\_units                 &  & 1                                              & Inv-NH3-bin (combine with on)   \\
NH3\_syn      & candidate\_units                 &  & 10                                             & Inv-NH3-cont (combine with on)  \\
NH3\_syn      & fom\_cost                        &  & 211646                                         & Inv-NH3-on                      \\
NH3\_syn      & is\_active                       &  & true                                           & Inv-NH3-on                      \\
NH3\_syn      & number\_of\_units                &  & 0                                              & Inv-NH3-on                      \\
NH3\_syn      & unit\_investment\_cost           &  & 626666                                         & Inv-NH3-on                      \\
NH3\_syn      & unit\_investment\_variable\_type &  & unit\_investment\_variable\_type\_integer      & Inv-NH3-bin (combine with on)   \\
NH3\_syn      & unit\_investment\_variable\_type &  & "unit\_investment\_variable\_type\_continuous" & Inv-NH3-cont (combine with on)  \\
NH3\_syn\_05  & candidate\_units                 &  & 1                                              & Inv-NH3-bin (combine with on)   \\
NH3\_syn\_05  & fom\_cost                        &  & 1058230                                        & Inv-NH3-bin (combine with on)   \\
NH3\_syn\_05  & is\_active                       &  & true                                           & Inv-NH3-bin (combine with on)   \\
NH3\_syn\_05  & number\_of\_units                &  & 0                                              & Inv-NH3-bin (combine with on)   \\
NH3\_syn\_05  & unit\_investment\_cost           &  & 3133329                                        & Inv-NH3-bin (combine with on)   \\
NH3\_syn\_05  & unit\_investment\_variable\_type &  & unit\_investment\_variable\_type\_integer      & Inv-NH3-bin (combine with on)   \\
... & & & \\
NH3\_syn\_10  & candidate\_units                 &  & 1                                              & Inv-NH3-bin (combine with on)   \\
NH3\_syn\_10  & fom\_cost                        &  & 2116460                                        & Inv-NH3-bin (combine with on)   \\
NH3\_syn\_10  & is\_active                       &  & true                                           & Inv-NH3-bin (combine with on)   \\
NH3\_syn\_10  & number\_of\_units                &  & 0                                              & Inv-NH3-bin (combine with on)   \\
NH3\_syn\_10  & unit\_investment\_cost           &  & 6266657                                        & Inv-NH3-bin (combine with on)   \\
NH3\_syn\_10  & unit\_investment\_variable\_type &  & unit\_investment\_variable\_type\_integer      & Inv-NH3-bin (combine with on)  \\ \hline
\end{tabular}
    }
\end{center}
\caption{Unit parameters.}
\label{tab:unit_param}
\end{table*}

\begin{table*}[h]
\begin{center} 
\scalebox{0.9}{
\begin{tabular}{|l|l|l|l|l|}
\hline
Node             & Parameter                           & Time                & Value                      & Scenario             \\ \hline
BESS             & candidate\_storages                 &                     & 10                         & Inv-storage-compress \\
BESS             & fix\_node\_state                    & 2018-12-31T23:00:00 & 0                          & Base                 \\
BESS             & frac\_state\_loss                   &                     & 4.00E-05                   & Base                 \\
BESS             & has\_state                          &                     & TRUE                       & Base                 \\
BESS             & node\_state\_cap                    &                     & 1                          & Inv-storage-compress \\
BESS             & storage\_investment\_cost           &                     & 13302                      & Inv-storage-compress \\
BESS             & storage\_investment\_variable\_type &                     & variable\_type\_continuous & Inv-storage-compress \\
El\_BESS         & nodal\_balance\_sense               &                     & ==                         & Base                 \\
El\_curtailed    & nodal\_balance\_sense               &                     & \textgreater{}=            & Base                 \\
El\_renew        & nodal\_balance\_sense               &                     & ==                         & Base                 \\
Group\_El        & nodal\_balance\_sense               &                     & ==                         & Base                 \\
Group\_H2\_mFRR  & nodal\_balance\_sense               &                     & ==                         & Base                 \\
H2               & nodal\_balance\_sense               &                     & ==                         & Base                 \\
H2\_decompressed & nodal\_balance\_sense               &                     & ==                         & Base                 \\
H2\_out          & nodal\_balance\_sense               &                     & \textgreater{}=            & Base                 \\
H2\_storage      & candidate\_storages                 &                     & 5                          & Inv-storage-compress \\
H2\_storage      & fix\_node\_state                    & 2018-12-31T23:00:00 & 0                          & Base                 \\
H2\_storage      & frac\_state\_loss                   &                     & 0.01                       & Base                 \\
H2\_storage      & has\_state                          &                     & TRUE                       & Base                 \\
H2\_storage      & node\_state\_cap                    &                     & 0.5                        & Inv-storage-compress \\
H2\_storage      & storage\_investment\_cost           &                     & 24229                      & Inv-storage-compress \\
H2\_storage      & storage\_investment\_variable\_type &                     & variable\_type\_integer    & Inv-storage-compress \\
H2O\_in          & nodal\_balance\_sense               &                     & \textless{}=               & Base                 \\
H2O\_out         & nodal\_balance\_sense               &                     & \textgreater{}=            & Base                 \\
Heat\_out        & nodal\_balance\_sense               &                     & \textgreater{}=            & Base                 \\
O\_out           & nodal\_balance\_sense               &                     & \textgreater{}=            & Base                 \\
Steam\_out       & nodal\_balance\_sense               &                     & \textgreater{}=            & Base               \\ \hline
\end{tabular}
    }
\end{center}
\caption{Node parameters.}
\label{tab:node_param}
\end{table*}

\begin{table*}[h]
\begin{center} 
\scalebox{0.7}{
\begin{tabular}{|l|l|l|l|l|l|l|l|}
\hline
\textbf{Parameter}               & \textbf{Relationship}                                                   & \textbf{Direction}   & \textbf{Unit}   & \textbf{Node}   & \textbf{Time} & \textbf{Value}     & \textbf{Scenario}                       \\ \hline
fix\_ratio\_in\_out\_unit\_flow  & unit\_\_node\_\_node\_BESS\_charge\_\_Group\_El\_\_BESS                 & unit\_\_node\_\_node & BESS\_charge    & Group\_El       &        & 1.015    & Base                                    \\
fix\_ratio\_in\_out\_unit\_flow  & unit\_\_node\_\_node\_BESS\_discharge\_\_BESS\_\_El\_BESS               & unit\_\_node\_\_node & BESS\_discharge & BESS            &        & 1.026    & Base                                    \\
fix\_ratio\_in\_out\_unit\_flow  & unit\_\_node\_\_node\_El\_curtail\_\_El\_renew\_\_El\_curtailed         & unit\_\_node\_\_node & El\_curtail     & El\_renew       &        & 1        & Base                                    \\
fix\_ratio\_in\_out\_unit\_flow  & unit\_\_node\_\_node\_H2\_compress\_\_Group\_El\_\_H2\_storage          & unit\_\_node\_\_node & H2\_compress    & Group\_El       &        & 4        & Base                                    \\
fix\_ratio\_in\_out\_unit\_flow  & unit\_\_node\_\_node\_H2\_compress\_\_H2\_\_H2\_storage                 & unit\_\_node\_\_node & H2\_compress    & H2              &        & 1        & Base                                    \\
fix\_ratio\_in\_out\_unit\_flow  & unit\_\_node\_\_node\_H2\_decompress\_\_H2\_storage\_\_H2\_decompressed & unit\_\_node\_\_node & H2\_decompress  & H2\_storage     &        & 1        & Base                                    \\
fix\_ratio\_in\_out\_unit\_flow  & unit\_\_node\_\_node\_H2\_discharge\_\_H2\_storage\_\_H2\_out           & unit\_\_node\_\_node & H2\_discharge   & H2\_storage     &        & 1        & Base                                    \\
fix\_ratio\_in\_out\_unit\_flow  & unit\_\_node\_\_node\_H2\_syn\_\_Group\_El\_\_Group\_H2\_mFRR           & unit\_\_node\_\_node & H2\_syn         & Group\_El       &        & 53.6     & Base                                    \\
fix\_ratio\_in\_out\_unit\_flow  & unit\_\_node\_\_node\_H2\_syn\_\_H2O\_in\_\_H2                          & unit\_\_node\_\_node & H2\_syn         & H2O\_in         &        & 9.9999   & Base                                    \\
fix\_ratio\_out\_out\_unit\_flow & unit\_\_node\_\_node\_H2\_syn\_\_H2\_\_Heat\_out                        & unit\_\_node\_\_node & H2\_syn         & H2              &        & 9.763    & Base                                    \\
fix\_ratio\_out\_out\_unit\_flow & unit\_\_node\_\_node\_H2\_syn\_\_H2\_\_O\_out                           & unit\_\_node\_\_node & H2\_syn         & H2              &        & 4.965    & Base                                    \\
fuel\_cost                       & unit\_\_from\_node\_BESS\_charge\_\_El\_renew                           & unit\_\_from\_node   & BESS\_charge    & El\_renew       & hourly & DK1 2019 & Base                                    \\
fuel\_cost                       & unit\_\_from\_node\_H2\_compress\_\_El\_renew                           & unit\_\_from\_node   & H2\_compress    & El\_renew       & hourly & DK1 2019 & Base                                    \\
fuel\_cost                       & unit\_\_from\_node\_H2\_syn\_\_El\_renew                                & unit\_\_from\_node   & H2\_syn         & El\_renew       & hourly & DK1 2019 & Base                                    \\
fuel\_cost                       & unit\_\_from\_node\_H2\_syn\_\_H2O\_in                                  & unit\_\_from\_node   & H2\_syn         & H2O\_in         &        & 10       & Base                                    \\
fuel\_cost                       & unit\_\_to\_node\_El\_curtail\_\_El\_curtailed                          & unit\_\_to\_node     & El\_curtail     & El\_curtailed   &        & -0.0001  & Base                                    \\
fuel\_cost                       & unit\_\_to\_node\_H2\_discharge\_\_H2\_out                              & unit\_\_to\_node     & H2\_discharge   & H2\_out         &        & -2160    & Base                                    \\
fuel\_cost                       & unit\_\_to\_node\_H2\_discharge\_\_H2\_out                              & unit\_\_to\_node     & H2\_discharge   & H2\_out         &        & -3240    & Premium-1.5x                            \\
fuel\_cost                       & unit\_\_to\_node\_H2\_discharge\_\_H2\_out                              & unit\_\_to\_node     & H2\_discharge   & H2\_out         &        & -4320    & Premium-2x                              \\
fuel\_cost                       & unit\_\_to\_node\_H2\_syn\_\_Heat\_out                                  & unit\_\_to\_node     & H2\_syn         & Heat\_out       &        & -2       & Base                                    \\
fuel\_cost                       & unit\_\_to\_node\_H2\_syn\_\_O\_out                                     & unit\_\_to\_node     & H2\_syn         & O\_out          &        & -1       & Base                                    \\
minimum\_operating\_point        & unit\_\_to\_node\_H2\_syn\_\_Group\_H2\_mFRR                            & unit\_\_to\_node     & H2\_syn         & Group\_H2\_mFRR &        & 0.2      & Mod-UC                                  \\
minimum\_operating\_point        & unit\_\_to\_node\_H2\_syn\_\_H2                                         & unit\_\_to\_node     & H2\_syn         & H2              &        & 0.2      & Mod-UC                                  \\
unit\_capacity                   & unit\_\_from\_node\_BESS\_discharge\_\_BESS                             & unit\_\_from\_node   & BESS\_discharge & BESS            &        & 1        & Base                                    \\
unit\_capacity                   & unit\_\_from\_node\_H2\_compress\_\_H2                                  & unit\_\_from\_node   & H2\_compress    & H2              &        & 0.06     & Base                                    \\
unit\_capacity                   & unit\_\_from\_node\_H2\_compress\_\_H2                                  & unit\_\_from\_node   & H2\_compress    & H2              &        & 1        & Inv-storage-compress                    \\
unit\_capacity                   & unit\_\_to\_node\_BESS\_charge\_\_BESS                                  & unit\_\_to\_node     & BESS\_charge    & BESS            &        & 1        & Base                                    \\
unit\_capacity                   & unit\_\_to\_node\_El\_PV\_\_El\_renew                                   & unit\_\_to\_node     & El\_PV          & El\_renew       &        & 27       & Base                                    \\
unit\_capacity                   & unit\_\_to\_node\_El\_wind\_\_El\_renew                                 & unit\_\_to\_node     & El\_wind        & El\_renew       &        & 54       & Base                                    \\
unit\_capacity                   & unit\_\_to\_node\_H2\_syn\_\_Group\_H2\_mFRR                            & unit\_\_to\_node     & H2\_syn         & Group\_H2\_mFRR &        & 0.235    & Base                                    \\
unit\_capacity                   & unit\_\_to\_node\_H2\_syn\_\_Group\_H2\_mFRR                            & unit\_\_to\_node     & H2\_syn         & Group\_H2\_mFRR &        & 0.0081   & Tech-H2-units (place before H2 invests) \\
unit\_capacity                   & unit\_\_to\_node\_H2\_syn\_\_H2                                         & unit\_\_to\_node     & H2\_syn         & H2              &        & 0.235    & Base                                    \\
unit\_capacity                   & unit\_\_to\_node\_H2\_syn\_\_H2                                         & unit\_\_to\_node     & H2\_syn         & H2              &        & 0.0081   & Tech-H2-units (place before H2 invests) \\ \hline
\end{tabular}
    }
\end{center}
\caption{Relationship parameters.}
\label{tab:node_param}
\end{table*}

\end{document}